\documentclass[11pt]{article}%

\usepackage{OCParxiv}

\usepackage[utf8]{inputenc} %
\usepackage[T1]{fontenc}    %
\usepackage[hidelinks]{hyperref}       %
\usepackage{url}            %
\usepackage{booktabs}       %
\usepackage{amsmath}
\usepackage{amsthm}
\usepackage{amssymb}
\usepackage{amsfonts}       %
\usepackage{nicefrac}       %
\usepackage{microtype}      %
\usepackage{tikz}
\usepackage{graphicx}
\usepackage[round]{natbib}
\usepackage{doi}
\usepackage[font=small]{caption}

\usepackage{numprint}%
\npdecimalsign{.}
\npthousandsep{,}

\usepackage{dsfont} %

\usepackage{color} %

\usepackage{algorithm}
\usepackage[noend]{algpseudocode}

\usetikzlibrary{arrows,shapes,trees,shapes.geometric}

\graphicspath{{figures/}}

\theoremstyle{plain}

\theoremstyle{remark}

\newcommand{\R}{\mathbb{R}}

\newcommand{\N}{\mathbb{N}}

\newcommand{\E}{\mathbb{E}}
\newcommand{\1}{\mathds{1}}
\newcommand{\bm}[1]{\mathbf{#1}}

\newcommand{\abs}[1]{\left\lvert#1\right\rvert}

\DeclareMathOperator*{\argmin}{arg\,min}

\title{Neural Networks for Extreme Quantile Regression\\with an Application to Forecasting of Flood Risk}

\date{} 					%

\author{Olivier~C.~Pasche\\
    Research Center for Statistics,\\
    University of Geneva, Switzerland,\\
    \href{mailto:olivier.pasche@unige.ch}{\texttt{olivier.pasche@unige.ch}}\\
    \And
    Sebastian~Engelke\\
    Research Center for Statistics,\\
    University of Geneva, Switzerland,\\
    \href{mailto:sebastian.engelke@unige.ch}{\texttt{sebastian.engelke@unige.ch}}\\
}

\renewcommand{\headeright}{}
\renewcommand{\undertitle}{}%
\renewcommand{\shorttitle}{Extreme Quantile Regression Neural Networks}

\hypersetup{
pdftitle={Neural Networks for Extreme Quantile Regression with an Application to Forecasting of Flood Risk},
pdfsubject={stat.ME},
pdfauthor={Olivier C. Pasche, Sebastian Engelke},
pdfkeywords={Extreme value theory, Generalized Pareto distribution, Machine learning, Prediction, Recurrent neural network},
}

\begin{document}
\maketitle

\begin{abstract}
Risk assessment for extreme events requires accurate estimation of high quantiles that go beyond the range of historical observations. When the risk depends on the values of observed predictors, regression techniques are used to interpolate in the predictor space. We propose the EQRN model that combines tools from neural networks and extreme value theory into a method capable of extrapolation in the presence of complex predictor dependence. Neural networks can naturally incorporate additional structure in the data. We develop a recurrent version of EQRN that is able to capture complex sequential dependence in time series. We apply this method to forecast flood risk in the Swiss Aare catchment. It exploits information from multiple covariates in space and time to provide one-day-ahead predictions of return levels and exceedance probabilities. This output complements the static return level from a traditional extreme value analysis, and the predictions are able to adapt to distributional shifts as experienced in a changing climate. Our model can help authorities to manage flooding more effectively and to minimize their disastrous impacts through early warning systems.
\end{abstract}

\keywords{Extreme value theory \and Generalized Pareto distribution \and Machine learning \and Prediction \and Recurrent neural network}

\section{Introduction}\label{s:intro}

Risk assessment is concerned with the analysis of rare events, which have small occurrence probabilities but carry the potential of serious impacts on our health, the environment, or the economy. Examples of such extreme events are floods in hydrology, crises in the financial system, or heatwaves in a changing climate.
In these applications, the quantity of interest is typically a univariate response variable $Y$ representing the random risk factor. The goal is to estimate a quantile $Q(\tau) = F^{-1}_Y(\tau)$ at level $\tau \in [0,1]$, where we denote by $F^{-1}_Y$ the generalized inverse of the distribution function of $Y$.

Since for risk quantification the level $\tau$ is usually very close to 1 so that the quantile $Q(\tau)$ goes beyond the range of the data, the classical approach is to model the tail of the distribution of $Y$ using extrapolation results from extreme value theory. Two main approaches exist. When $Y$ represents, say, a daily quantity, then the generalized Pareto distribution (GPD) can be used to approximate the tail above a high threshold $u$ by~\citep{GPDBalkema,GPDPickands}
\begin{align}\label{GPD}
\mathbb P(Y > y) \approx \mathbb P(Y > u) \left(1 + \xi \frac{y-u}{\sigma(u)}\right)_+^{-1/\xi},\quad y \geq u,
\end{align}
where $\xi \in \mathbb R$ and $\sigma(u)>0$ are the shape and scale parameters, and the second factor on the right-hand side is the GPD approximation to $\mathbb P(Y>y \mid Y>u)$.  On the other hand, if $Y$ represents an annual maximum, then the generalized extreme value distribution provides a good fit \citep{fis1928}.
In hydrology and climate science, risk is often assessed as the $T$-year return level $Q^T$, that is, the size of an event that is exceeded on average once every $T$ years. If $Y$ represent a quantity with $n_Y$ independent recordings per year (e.g., $n_Y=365$ for daily data and $n_Y=1$ for annual maxima), then the $T$-year return level is the quantile $Q^T = Q(1 - 1/(n_Y T))$.

\begin{figure}[t]
\centering
\includegraphics[width=0.85\textwidth]{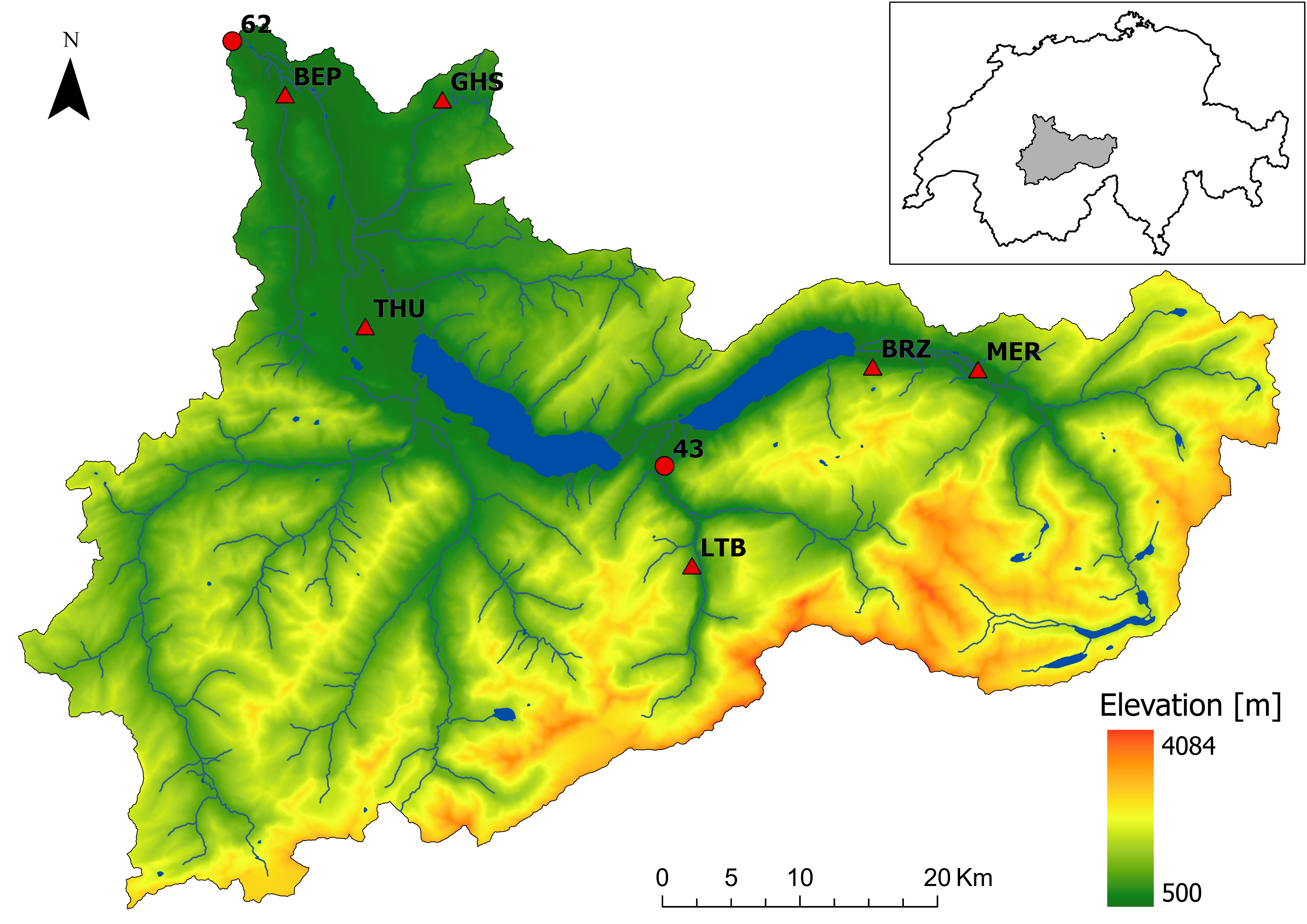}%
\caption{Topographic map of water catchment of the gauging station in Bern--Sch\"onau (62) on the Aare in Switzerland. Another gauging station upstream in Gsteig (43) on the L\"utschine river and six meteorological stations with precipitation measurements (triangles) are also shown.}%
\label{f:chstations}
\end{figure}

\begin{figure}[t]%
\centering
\includegraphics[width=0.9\textwidth]{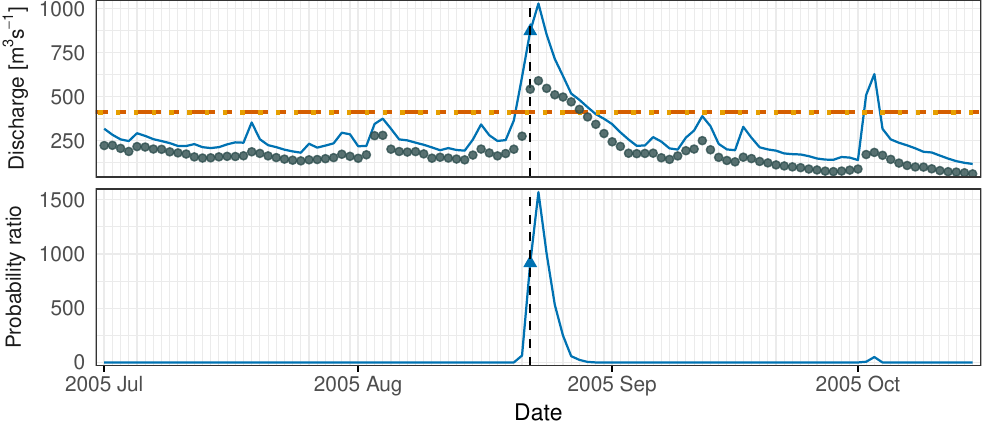}%
\caption{Top: Daily average discharge (points) at the Bern--Sch\"onau station (62) 
and one-day-ahead EQRN forecasts of conditional 100-year quantiles (solid line) during the 2005 flood. Horizontal lines show unconditional $Q^{100}$ based on GEV (dashed) and GPD (dotted). Bottom: One-day-ahead EQRN forecast of the conditional probability of exceeding the GEV estimated $Q^{100}$ as a ratio to the unconditional probability. The vertical line indicates August 22, the day of the first exceedance.}%
\label{f:chpreds2005}%
\end{figure}

Figure~\ref{f:chstations} shows the river catchment of a gauging station on the Aare river in Bern, Switzerland. 
It is part of the Aare--Rhein basin, where flooding is a major economic and safety concern \citep{EXAR}.
The Swiss Federal Office for the Environment (FOEN) provides recordings of daily average discharges throughout the country. For risk assessment they report the 100-year return levels using the GEV method for annual maxima and the GPD for large daily discharges. The horizontal lines in Figure~\ref{f:chpreds2005} show estimates of this return level at the Bernese station based on data from the years 1930--1958. Both methods give very similar results.

The disadvantage of such an unconditional approach is twofold. First, the return level is static and unable to reflect changes in the size of extreme floods over time, which can occur, for instance, due to climate change. For the Bernese station, for instance, a structural break has been observed in the nineties without a clearly defined cause\footnote{See flood report of the FOEN at \href{https://www.hydrodaten.admin.ch/en/2135.html}{\texttt{https://www.hydrodaten.admin.ch/en/2135.html}}.}, making classical extreme value modelling challenging.
Second, while the return level $Q^T$ is relevant for the construction of long-term flood infrastructure, it can not be used to assess the risk of flooding on a given day. Such forecasting of extreme events is crucial for early warning systems. Indeed, the probability of exceeding on a particular day a given high threshold, say the (constant) 100-year return level, depends on many covariates $\bm X$ such as the river flows upstream and precipitation in the catchments during the preceding days and weeks.

In this paper, we, therefore, advocate a conditional version of return levels defined as the conditional quantile of $Y$ given a vector of observed covariates $\bm X = \bm x$, that is, 
\begin{equation}\label{e:condquantile}%
Q_{\bm x}(\tau) =F^{-1}_{Y\mid \bm{X} = \bm{x}}(\tau).
\end{equation}%
The interpretation of such a conditional $T$-year return level $Q^T_{\bm x} = Q_{\bm x}(1-1/(n_Y T))$ is different from the unconditional return level $Q^T$. 
Since $Q^T_{\bm x}$ depends on the exact configuration of the covariates $\bm x$, one can see a conditional $T$-year return level as the size of an event that is exceeded in average once every $T$ years, if the covariate vector $\bm X$ of all observations of $Y$ had the same value $\bm x$.
A more precise interpretation is to see $Q^T_{\bm x}$ as the value that is exceeded in the next time step with probability $1/(n_Y T)$, and we refer to it as the conditional quantile with return period $T$ years; for a comprehensive discussion of return levels and quantiles, see~\citet{BMvsPOT}. 

The top panel of Figure~\ref{f:chpreds2005} shows one-day-ahead forecasts of such conditional 100-year quantiles $Q^{100}_{\bm x}$ for the Bernese river data from the method of this paper, fitted to the years 1930--1958. We see that, as opposed to the unconditional return level $Q^{100}$, the conditional quantile changes from day to day, depending on past precipitation and river flows.
In fact, on August 21, the day before the first exceedance of the unconditional $Q^{100}$ during the serious flood of August 2005 in Bern, the size of the conditional 100-year event predicted for the next day was much higher than on other days, and the forecasted conditional probability of such an exceedance (bottom panel of Figure~\ref{f:chpreds2005}) was 920 times larger than the static 100-year probability. 
Both outputs can be used as triggers for early warnings and additional flood management measures. In particular, the days when an exceedance above $Q^{100}$ is likely can be effectively pinpointed thanks to their temporal sparsity in the probability forecast.

Forecasting extreme events is notoriously difficult due to the low occurrence probabilities involved. This is particularly true for hydrological models, which are typically used by national agencies and which do not use explicit tail extrapolation. In the aftermath of the 2005 flood, the FOEN published a detailed analysis of the internal forecasting procedures during this event~\citep{BAFU2005floodT1}. The report showed that too-late warnings lead to more severe consequences since forecasters did not trust the predicted precipitation amounts during this extreme scenario and underestimated the flood risk. This gives another motivation why our statistical model and its output in Figure~\ref{f:chpreds2005} could have been a helpful tool during this flood.

Conditional quantiles~\eqref{e:condquantile} are studied in the field of quantile regression, where many flexible methods exist~\citep{QRF, QRNN,iQRNN,GRF}.
While they work well for quantile levels within the data range, they break down for extreme values of $\tau$ close to 1. Such extreme quantile regression relies on extrapolation results as in~\eqref{GPD}, where extreme value parameters, such as the scale $\sigma(\bm x)$ and shape $\xi(\bm x)$, may be modelled as functions of the covariates through linear models \citep{Wang2012, EXQAR}, generalized additive models~\citep{ExGAM,ExGAM2}, or kernel methods \citep{abdelaati2011, GardesStupfler2019, Velthoenetal2019}. To overcome limitations of additive and kernel-based methods in higher dimensions, more recently, flexible tree-based methods have been combined with the GPD extrapolation for predicting extreme conditional quantiles~\citep{gbex,erf} or predictive tail distributions~\citep{Kohfireboosting} on complex data. %
Tree-based methods have the advantage of requiring little tuning for good prediction performance. However, they can not incorporate additional structure of the data as encountered in time series or spatial applications.

The goal of our work is to combine ideas from extreme value theory and machine learning to propose an extreme quantile regression model that has the ability to extrapolate in the direction of the response $Y$ and to model complex covariate dependencies in the predictors $\bm X$. We propose an extreme quantile regression network (EQRN) that models covariate-dependent GPD parameters $\sigma(\bm{x})$ and $\xi(\bm{x})$ as outputs of a neural network. Conditional quantile estimates at the desired extreme level are then readily derived from the estimated conditional tail distribution.
Neural networks are known for their ability to model complex dependencies and to approximate any measurable function arbitrarily well~\citep{NNapprox2}.
The second advantage is versatility. The deep learning literature is rich in network architectures, activation functions, and regularization methods. In particular, convolutions produce shift-invariant models for covariates with spatial dependencies, and recurrent architectures provide models for sequentially dependent observations such as time series. 
In our application of flood forecasting with time-dependent data, recurrent neural networks~\citep{RNN1,RNN2} are of particular interest.

The main output of our EQRN model for sequentially dependent data is the one-day-ahead risk forecast, either in terms of conditional quantiles or exceedance probabilities. Thanks to the GPD approximation, the method can extrapolate beyond the range of the data, as illustrated in Figure~\ref{f:chpreds2005}. The strength of our recurrent model lies in the ability to exploit information from multiple covariates and capture complex time dependence. It is, therefore, an effective early warning tool even for unprecedented, record-shattering events, like the 2005 floods. This is of particular importance in a nonstationary system, where climate change makes extreme events increasingly likely \citep{fis2021}.

The main contributions of the paper are threefold.
First, our EQRN method is the first to model the conditional GPD parameters
through neural networks. Thanks to the large number of neural network architectures, this expands the range of possible applications for extreme quantile regression to new areas. While we concentrate on sequentially dependent data, our method can be used, for instance, in combination with convolutional or graphical neural networks for spatial covariates. Second, a major technical contribution is to make neural networks applicable in the extreme value context. To stabilize the prediction of extreme quantiles in these highly flexible regression
methods, we propose to use an orthogonal reparametrization of the GPD deviance, a suitable choice of activation
functions, and the use of the intermediate quantiles as additional covariates; the latter is a new idea that seems to also improve other extreme quantile methods. Finally, a main novelty is the application to flood forecasting and the notion of conditional return levels that can be used as early warnings. With this perspective our method enables applications in many other areas, such as the one-day-ahead forecasting of the value-at-risk or of the expected shortfall in financial time series.

The paper is organized as follows. In Section~\ref{s:background} we provide background on quantile regression, extreme value theory, and neural networks. We propose our EQRN model for both independent and sequentially dependent data in Section~\ref{s:eqrnn}. Section~\ref{s:sims} contains a simulation study to assess the performance of our approach in comparison to existing methods. In Section~\ref{s:appl} we describe the Swiss river data, apply our methodology to forecast flood risk and discuss the implications of the results. Section~\ref{s:conclu} concludes with a brief discussion.

\section{Background}\label{s:background}

\subsection{Quantile regression}\label{ss:qr}

In the classical quantile regression setup, we observe an independent and identically distributed sample $\mathcal{D}=\left\{(\bm{x}_i, y_i)\right\}_{i=1}^n$ of the random vector $(\bm{X},Y)$, where $Y$ is the real-valued response variable and $\bm{X}$ is a vector of $p$ covariates (or predictors). One aims at predicting the conditional quantile $Q_{\bm x}(\tau)$ defined in~\eqref{e:condquantile} of $Y$, given $\bm{X} = \bm{x}$, for some predictor value $\bm{x}\in\R^p$ and probability level $\tau \in (0, 1)$ of interest.

Analogously to regression that minimizes the mean squared error, quantile regression minimizes the {quantile loss},%
\begin{equation}\label{qloss}
Q_{\bm{x}}(\tau) = \argmin_{q} \E[\rho_{\tau}(Y - q) \mid \bm{X} = \bm{x}],
\end{equation}
where $\rho_{\tau}(t) := t(\tau-\1_{\{ t<0\}})$ is the {quantile check function} \citep{koen1978}. Many parametric and nonparametric quantile regression models exist, including linear models~\citep{chernozhukov2005}, random forests~\citep{GRF}, and neural networks~\citep{QRNN,iQRNN}. They yield a conditional quantile estimate by minimizing the empirical quantile loss over the training sample $\mathcal{D}$, that is,
\begin{equation}\label{emp_qloss}
\hat{Q}_{\bm{x}}(\tau) = \argmin_{q_\tau\in\mathcal{M}} \dfrac{1}{n}\sum_{i=1}^n \rho_{\tau}(y_i - q_\tau (\bm{x}_i)),
\end{equation}
where $\mathcal{M}$ is the set of possible quantile functions $q_\tau(\cdot)$ characterized by the model.

Classical methods for quantile regression that rely on the quantile loss~\eqref{qloss} perform well for ``moderate'' probability levels $\tau$. To define what that means exactly, we typically let $\tau_n$ depend on the sample size $n$. The expected number of exceedances of $y_i$  over the respective conditional quantile $Q_{\bm x_i}(\tau_n)$, $i=1,\dots, n$, is given by $n(1-\tau_n)$.
With moderately extreme, or intermediate, we refer to a sequence $\tau_n \to 1$ with $n(1-\tau_n) \to \infty$, meaning that the quantile goes to the upper endpoint of the distribution, but there are more and more exceedances with growing sample size $n$.
On the other hand, we call a quantile level $\tau_n\to 1$ extreme if $n(1-\tau_n) \to c \in [0,\infty)$; that is, there are finitely many, or possibly zero, exceedances over  $Q_{\bm x_i}(\tau_n)$ in the sample. In this situation, classical quantile regression methods do not perform well due to the scarcity of observations in the tail of the response.

The left panel of Figure~\ref{f:motivMLP} illustrates this issue for a sample $y_1,\dots, y_n$ with $n=1000$ and no covariates. The dashed line shows for different quantile levels $\tau_n$ the empirical quantile estimates, obtained by solving the respective quantile loss function without covariates. It can be seen that as soon as the number of exceedances $n(1-\tau_n) < 1$, that is, $\tau_n > 99.9\%$, there is a significant bias compared to the true quantiles (solid line). The reason is that the empirical estimates can not predict higher than the largest observation. When covariates are present, this issue persists, since quantile regression relies on solving the empirical quantile loss~\eqref{emp_qloss}.

In the sequel we omit the dependence on $n$ and write $\tau$ instead of $\tau_n$. Intermediate quantile levels will be denoted by $\tau_0$.

\subsection{Generalized Pareto distribution}\label{ss:pot}

In order to predict well on extreme quantiles, a method should rely on asymptotic results from extreme value theory for accurate extrapolation beyond the sample. In particular, we rely on the generalized Pareto distribution (GPD).
In the presence of covariates, it arises as an approximation of the tail of the distribution of $Y \mid \bm X = \bm x$. More precisely, we use a conditional version of~\eqref{GPD},
\begin{align}\label{GPD_cond}
\mathbb P(Y > y\mid \bm X = \bm x) \approx (1-\tau_0) \left(1 + \xi(\bm x) \frac{y-u(\bm x)}{\sigma(\bm x)}\right)_+^{-1/\xi(\bm x)}, \quad y > u(\bm x),
\end{align}
where the threshold $u(\bm x)$ is chosen as an intermediate quantile $Q_{\bm x}(\tau_0)$ at level $\tau_0\in (0,1)$ close to 1, and the shape $\xi(\bm x) \in \mathbb R$ and scale $\sigma(\bm x)>0$ depend on the covariates; here we omit the dependence of $\sigma(\bm x)$ on the intermediate level $\tau_0$ in the notation. This approximation holds under weak conditions on the tail of $Y \mid \bm X = \bm x$; see \citet{GPDBalkema,GPDPickands} for the precise statement. This condition is of univariate nature, and it can, therefore, be verified even in more complex situations, for instance, where $\bm X$ represents the history of a multivariate time series; see Section~\ref{ss:simssetup} for details.
The shape parameter $\xi(\bm x)$ is important since it encodes the tail heaviness of the response: if it is positive, the response has a heavy-tailed distribution such as Pareto or Student-$t$; if it is zero, the response is light-tailed such as a Gaussian or exponential; if it is negative, then the response has a finite upper endpoint.

In order to predict an extreme quantile at level $\tau > \tau_0$ from approximation~\eqref{GPD_cond}, we can invert this expression to find
\begin{equation}\label{e:gpdquant}
Q_{\bm x}(\tau) := Q_{\bm x}(\tau_0) + \frac{\sigma(\bm x)}{\xi(\bm x)} \left[\left(\dfrac{1 - \tau_0}{1-\tau}\right)^{\xi(\bm x)} -1 \right].%
\end{equation}
This shows that an estimate $\hat Q_{\bm x}(\tau)$ of an extreme quantile requires estimates of the intermediate quantile $\hat Q_{\bm x}(\tau_0)$ and of the conditional GPD parameters $\hat \xi(\bm x)$ and $\hat \sigma(\bm x)$ as functions of the predictor vector.
For the intermediate quantile function, we can use any of the existing methods for quantile regression since they work well for this moderate quantile level, as discussed above. Estimation of the GPD parameters can be done by specifying a parametric or nonparametric model. We will use neural networks for this purpose, which are introduced in the next section.

\begin{figure}[t]
\centering
\hfill\includegraphics[width=0.38\textwidth]{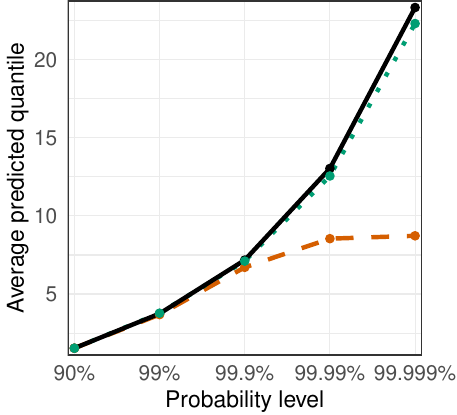}\hfill%
\begin{tikzpicture}[shorten >=1pt,->]
\definecolor{mcolr}{RGB}{213,94,0}
\definecolor{mcolb}{RGB}{0,114,178}
\definecolor{mcolg}{RGB}{0,158,115}
\definecolor{mcoly}{RGB}{230,159,0}

\def\stepsep{50pt}%
\def\cropsepb{50pt}
\def\cropsept{20pt}
\def\elabsep{9.375pt}
\tikzstyle{annot} = [text width=4em, text centered]
\tikzstyle{layer}=[circle, minimum size=25.0pt, fill=white, line width=1.25pt, draw=black, inner sep=1pt]
\tikzstyle{data}=[layer, fill=mcolb!75]
\tikzstyle{output}=[layer, fill=mcolg!75]
\tikzstyle{loss}=[regular polygon, regular polygon sides=4, minimum size=35.0pt, fill=mcolr!70, line width=1.25pt, draw=black]
\tikzstyle{fctarr}=[line width=1.25, ->, color=black]
\tikzstyle{usearr}=[line width=1.25, ->, dashed, color=black]
\tikzstyle{dot}=[mark size=1pt, color=black]
\tikzstyle{pholder}=[mark size=0pt, color=black]
\tikzstyle{lstm}=[regular polygon, regular polygon sides=4, minimum size=37.0pt, fill=mcoly!20, line width=1.25pt, draw=black, inner sep=0pt]

\node (x) [data] at (0pt, 0pt) {$\mathbf{x}^{}$};
\node (l1) [layer] at (1*\stepsep, 0pt) {$\mathbf{x}^{(1)}$};
\node (l2) [layer] at (2*\stepsep, 0pt) {$\mathbf{x}^{(2)}$};
\node (l3) [layer] at (3*\stepsep, 0pt) {$\mathbf{x}^{(L)}$};
\node (out) [output] at (4*\stepsep, 0pt) {$g_\mathcal{W}(\mathbf{x})$};

\node (y) [data] at (2*\stepsep, \stepsep) {$y$};
\node (L) [loss] at (3*\stepsep, \stepsep) {$\ell$};
\draw [fctarr] (x) to  (l1);
\draw [fctarr] (l1) to  (l2);
\node[dot] at (2.5*\stepsep-2mm,0) {\pgfuseplotmark{*}};
\node[dot] at (2.5*\stepsep,0) {\pgfuseplotmark{*}};
\node[dot] at (2.5*\stepsep+2mm,0) {\pgfuseplotmark{*}};
\draw [fctarr] (l3) to  (out);
\draw [usearr] (out) to [in=0, out=90] (L);
\draw [usearr] (y) to (L);
\node[pholder] at (2*\stepsep,-\cropsepb) {};
\node[pholder] at (2*\stepsep,\stepsep+\cropsept) {};
\end{tikzpicture}\hfill
\caption{Left: True $\tau$-quantiles (solid line) compared to empirical estimates (dashed line) and GPD based estimates (dotted line) for moderate to extreme probability levels (log-scale) for sample size \numprint{1000}. Estimates are averages over \numprint{100} trials. Right: Multilayer perceptron flowchart from input $\bm{x}$ to output $g_\mathcal{W}(\bm{x})$, with loss function $\ell$ and corresponding response $y$.}
\label{f:motivMLP}
\end{figure}
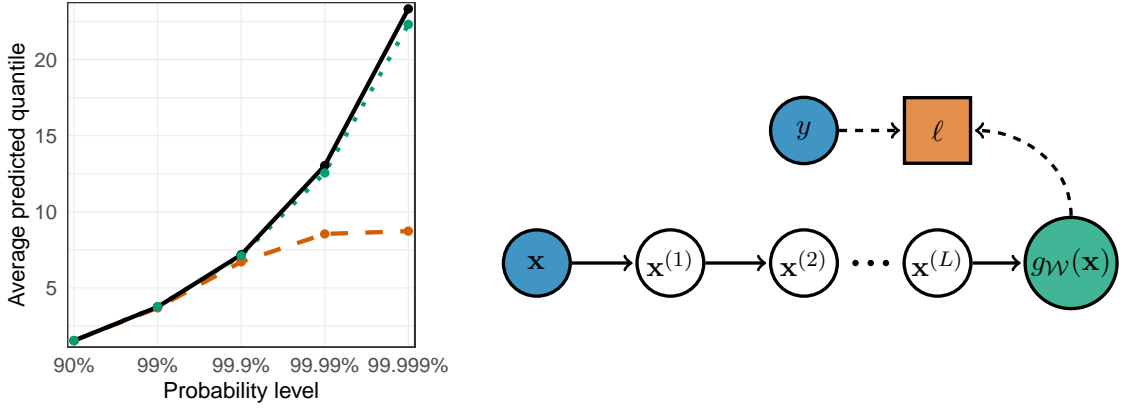

The green line in Figure~\ref{f:motivMLP} shows estimates $\hat Q(\tau)$ for different quantile levels $\tau$, using the approximation~\eqref{e:gpdquant} without covariate dependence, with empirical intermediate quantile at $\tau_0 = 90\%$ and GPD parameters estimated with maximum likelihood. It can be seen that the extrapolation solves the bias issue of empirical methods.

\subsection{Neural networks and conditional density estimation}\label{ss:cdenn}

The literature on neural networks is vast, and existing methods are being improved constantly. We concentrate in this section on well-established techniques that are most relevant for our purpose of modelling extreme quantiles.

A {multilayer perceptron (MLP)} or fully-connected feed-forward neural network model is a parametric family of nonlinear functions $g_\mathcal{W}: \R^p \rightarrow \R^q$ that map a $p$-dimensional input $\bm x$ to a $q$-dimensional output by 
\begin{equation}\label{e:mlp}
        \bm{x} \mapsto \bm{x}^{(L+1)},\; \text{ with }\; \bm{x}^{(l)}= \sigma^l\left(W^l\bm{x}^{(l-1)}+b^l\right) \; \forall l=1,\ldots,L+1,
\end{equation}
where $\bm{x}^{(0)} = \bm{x}$. The number of hidden layers $L\in\N$, the hidden layer dimensions $h_1,\ldots,h_L\in\N$, and the choice of activation functions $\sigma^l:\R^{h_l} \rightarrow \R^{h_l}, l=1,\ldots,L+1$ (with $h_0=p$ and $h_{L+1}=q$) are hyperparameters that need to be chosen, for instance, by cross-validation. The set of trainable parameters to be inferred from data contains all weights and bias terms of the network, that is, $\mathcal{W}=\left\{(W^l,b^l);l=1,\ldots,L+1 \right\}$, with $W^l\in\R^{h_{l}\times h_{l-1}}$ and $b^l\in\R^{h_l}$.
Figure~\ref{f:motivMLP} shows a schematic illustration of the transformations inside the MLP.

In the general setting, $p$ is the number of features or covariates considered in the model, and $q$ depends on the task at hand.
In order to train a model, a loss function
$\ell: \mathbb R \times \mathbb R \to [0,\infty)$ is required that maps a tuple $(y,g_\mathcal{W}(\bm x))$ of response and prediction to a positive number quantifying their discrepancy. Common tasks include mean regression with $q=1$ and squared error loss, quantile regression with $q=1$ and quantile loss, and classification with $q$ equal to the number of possible classes and cross-entropy as loss.

For conditional density estimation, or distribution regression, we suppose that $Y$ follows a distribution with parametric probability density $f_Y(\cdot ; {\theta})$ and parameter ${\theta}={\theta}(\bm{x})$ depending on the vector $\bm{X} = \bm x$. Conditional density estimation networks are neural networks that aim at outputting conditional estimates  $g_\mathcal{W}(\bm x) = {\theta}(\bm x)$ based on realizations of $\bm{X} = \bm x$ as input~\citep[e.g.,][]{CDENCannon}. In this setting, $p$ is the dimension of $\bm{X}$, and $q$ is the dimension of ${\theta}$. The loss function is the deviance or negative log-likelihood loss $\ell(y,{{\theta}}(\bm x))=-\log f_Y(y ; {{\theta}}(\bm x))$. 

To train a neural network on the training dataset $\mathcal{D}=\left\{(\bm{x}_i, y_i)\right\}_{i=1}^n$, we find the optimal parameter values minimizing the average empirical loss, that is,
\begin{equation}\label{e:nnoptim}
    \hat{\mathcal{W}}\in\argmin_{\mathcal{W}}\frac{1}{n}\sum_{i=1}^{n}\ell(y_i,g_\mathcal{W}(\bm{x}_i)).
\end{equation}
This is generally achieved via backpropagation using mini-batches and optimization algorithms, such as the well-performing gradient descent variants~\citep{Adam, RMSProp, AdaGrad}.
Since neural networks are typically overparameterized, overfitting has to be prevented with regularization methods, such as $L_2$ weight penalties for narrow networks and dropout~\citep{dropout} for deeper architectures.
As the optimization problem~\eqref{e:nnoptim} is often nonconvex, local-minima convergence is an issue.
Restricting training to a subset of $\mathcal{D}$ and keeping the rest to track the validation loss at the end of each epoch helps to avoid local minima by learning rate decay.
Restarting training with different initializations and keeping the best fit in terms of validation loss often leads to lower minima.
Early stopping based on the validation loss is another measure against overfitting.
The final validation loss is used for model selection and the choice of optimal hyperparameters; for more details on the fitting of neural networks, see \citet{DL}.

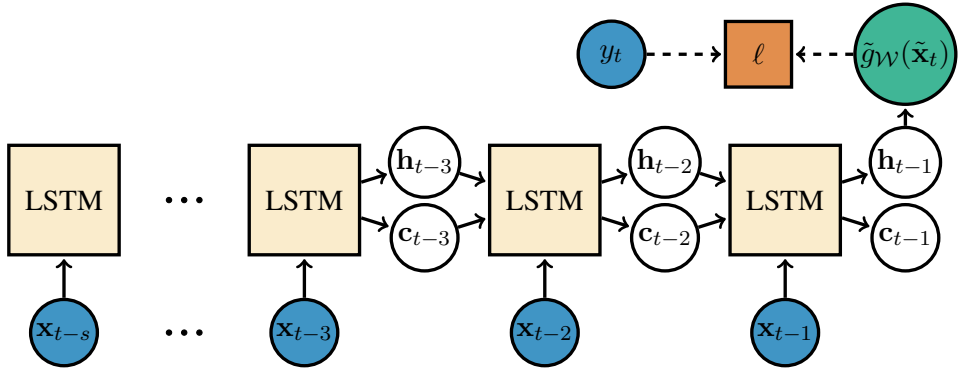
\begin{figure}[t]
\centering
\begin{tikzpicture}[shorten >=1pt,->]
\definecolor{mcolr}{RGB}{213,94,0}
\definecolor{mcolb}{RGB}{0,114,178}
\definecolor{mcolg}{RGB}{0,158,115}
\definecolor{mcoly}{RGB}{230,159,0}

\def\featsep{50pt}
\def\stepsep{55pt}
\def\elabsep{9.375pt}
\def\lstmsep{90pt}
\tikzstyle{annot} = [text width=4em, text centered]
\tikzstyle{layer}=[circle, minimum size=25.0pt, fill=white, line width=1.25pt, draw=black, inner sep=1pt]
\tikzstyle{data}=[layer, fill=mcolb!75]
\tikzstyle{output}=[layer, fill=mcolg!75]
\tikzstyle{loss}=[regular polygon, regular polygon sides=4, minimum size=35.0pt, fill=mcolr!70, line width=1.25pt, draw=black]
\tikzstyle{fctarr}=[line width=1.25, ->, color=black]
\tikzstyle{usearr}=[line width=1.25, ->, dashed, color=black]
\tikzstyle{dot}=[mark size=1pt, color=black]
\tikzstyle{lstm}=[regular polygon, regular polygon sides=4, minimum size=37.0pt, fill=mcoly!20, line width=1.25pt, draw=black, inner sep=0pt]

\node (xt) [data] at (0pt, 0pt) {$\mathbf{x}_{t-1}$};
\node (xt1) [data] at (-1*\lstmsep, 0pt) {$\mathbf{x}_{t-2}$};
\node (xt2) [data] at (-2*\lstmsep, 0pt) {$\mathbf{x}_{t-3}$};
\node (xts) [data] at (-3*\lstmsep, 0pt) {$\mathbf{x}_{t-s}$};

\node (lst) [lstm] at (0pt, \featsep) {LSTM};
\node (lst1) [lstm] at (-1*\lstmsep, \featsep) {LSTM};
\node (lst2) [lstm] at (-2*\lstmsep, \featsep) {LSTM};
\node (lsts) [lstm] at (-3*\lstmsep, \featsep) {LSTM};

\node (ht) [layer] at (0.5*\lstmsep, \featsep+5mm) {$\mathbf{h}_{t-1}$};
\node (ht1) [layer] at (-0.5*\lstmsep, \featsep+5mm) {$\mathbf{h}_{t-2}$};
\node (ht2) [layer] at (-1.5*\lstmsep, \featsep+5mm) {$\mathbf{h}_{t-3}$};
\node (ct) [layer] at (0.5*\lstmsep, \featsep-5mm) {$\mathbf{c}_{t-1}$};
\node (ct1) [layer] at (-0.5*\lstmsep, \featsep-5mm) {$\mathbf{c}_{t-2}$};
\node (ct2) [layer] at (-1.5*\lstmsep, \featsep-5mm) {$\mathbf{c}_{t-3}$};

\node (out) [output] at (0.5*\lstmsep, \featsep+\stepsep) {$\tilde{g}_\mathcal{W}(\tilde{\mathbf{x}}_{t})$};
\node (y) [data] at (0.5*\lstmsep-2*\stepsep, \featsep+\stepsep) {$y_{t}$};
\node (L) [loss] at (0.5*\lstmsep-1*\stepsep, \featsep+\stepsep) {$\ell$};

\draw [fctarr] (xt) to  (lst);
\draw [fctarr] (xt1) to  (lst1);
\draw [fctarr] (xt2) to  (lst2);
\draw [fctarr] (xts) to  (lsts);

\draw [fctarr] (lst) to  (ht);
\draw [fctarr] (lst1) to  (ht1);
\draw [fctarr] (ht1) to  (lst);
\draw [fctarr] (lst2) to  (ht2);
\draw [fctarr] (ht2) to  (lst1);
\draw [fctarr] (lst) to  (ct);
\draw [fctarr] (lst1) to  (ct1);
\draw [fctarr] (ct1) to  (lst);
\draw [fctarr] (lst2) to  (ct2);
\draw [fctarr] (ct2) to  (lst1);

\draw [fctarr] (ht) to  (out);

\node[dot] at (-2.5*\lstmsep-2mm,0) {\pgfuseplotmark{*}};
\node[dot] at (-2.5*\lstmsep,0) {\pgfuseplotmark{*}};
\node[dot] at (-2.5*\lstmsep+2mm,0) {\pgfuseplotmark{*}};
\node[dot] at (-2.5*\lstmsep-2mm,\featsep) {\pgfuseplotmark{*}};
\node[dot] at (-2.5*\lstmsep,\featsep) {\pgfuseplotmark{*}};
\node[dot] at (-2.5*\lstmsep+2mm,\featsep) {\pgfuseplotmark{*}};
\draw [usearr] (out) to (L);
\draw [usearr] (y) to (L);
\end{tikzpicture}
\caption{Single-layer LSTM network flowchart from input $\tilde{\bm{x}}_t:=(\bm{x}_{t-s},\ldots,\bm{x}_{t-1})$ to output $\tilde{g}_\mathcal{W}(\tilde{\bm{x}}_t)$, with loss evaluation. The LSTM cells represent the transformation in~\eqref{lstm_cell}.}%
\label{f:LSTMdiagr}
\end{figure}

When observations are dependent in space or time, generalizations of the MLP exist to account for these particular structures. 
Convolutional and graph neural networks exploit neighbourhood information with parsimonious architectures that are effective for images, graphs, or spatial observations~\citep{DLLeCun,GNN1}. 
We concentrate here on methods for sequential dependence that arises typically in time series $\left\{(\bm{X}_t, Y_t)\right\}_{t=1}^T$. For this type of data, recurrent architectures of the network allow capturing dependence between observations.
A simple recurrent neural network (RNN) layer~\citep{RNN1,RNN2} takes as input a vector $\bm{x}_t$ and outputs the hidden recurrent state vector
\begin{equation*}
\bm{h}_t = \tanh(W_{\rm xh} \bm{x}_t + W_{\rm hh} \bm{h}_{t-1} + \bm{b}_{\rm h}),
\end{equation*}
depending both on $\bm{x}_t$ and the hidden state $\bm{h}_{t-1}$ recursively resulting from the previous inputs $\bm{x}_{t-1}$ and $\bm{h}_{t-2}$ in the sequence. Here and in the sequel, the bias vectors $b_\cdot$ and weight matrices~$W_{\cdot \cdot}$, indexed by the input and output variables, are the trainable parameters.
This model has then been improved by the addition of a gating cell state $\bm{c}_t$, to avoid vanishing gradient issues, as well as a forget gate $\bm{f}_t$, an input gate $\bm{i}_t$, and output gate $\bm{o}_t$, to control both short- and long-term dependencies in the sequence. This yields the long short-term memory (LSTM) layer~\citep{LSTM1,LSTM2,LSTM3,LSTMGRUcomp}
\begin{equation}\label{lstm_cell}
\begin{array}{ll}
\bm{i}_t = \sigma(W_{\rm xi} \bm{x}_t + W_{\rm hi} \bm{h}_{t-1} + \bm{b}_{\rm i}), & \bm{f}_t = \sigma(W_{\rm xf} \bm{x}_t + W_{\rm hf} \bm{h}_{t-1} + \bm{b}_{\rm f}), \\
\bm{g}_t = \tanh(W_{\rm xg} \bm{x}_t + W_{\rm hg} \bm{h}_{t-1} + \bm{b}_{\rm g}), & \bm{o}_t = \sigma(W_{\rm xo} \bm{x}_t + W_{\rm ho} \bm{h}_{t-1} + \bm{b}_{\rm o}), \\
\bm{c}_t = \bm{f}_t \odot \bm{c}_{t-1} + \bm{i}_t \odot \bm{g}_t, & \bm{h}_t = \bm{o}_t \odot \tanh(\bm{c}_t),
\end{array}
\end{equation}
where $\sigma(z)=1 / (1 + \exp(-z))$ is the sigmoid activation and $\odot$ is the Hadamard (or componentwise) product; see Figure~\ref{f:LSTMdiagr} for an illustration.
The common dimension of the vectors defined in~\eqref{lstm_cell}, is a hyperparameter of the layer.
The input of this layer $\tilde{\bm{x}}_t:=(\bm{x}_{t-s},\ldots,\bm{x}_{t-1})$ can include predictors from the past to model longer dependencies, where $s \in \mathbb N$ determines the time horizon.

The LSTM model has been simplified by~\citet{GRU} into the gated recurrent unit (GRU) layer, which has become a popular alternative.
A multilayer recurrent network is obtained by considering the $\bm{h}_t$ as a sequence of inputs for the following recurrent layers; see Figure~\ref{f:MLLSTMdiagr} in Supplementary Material~\ref{ss:illus}. Usually, a fully connected layer as in~\eqref{e:mlp} is used to map the hidden state of the final recurrent layer to the network output $\tilde{g}_\mathcal{W}(\tilde{\bm{x}}_t)$.

\section{Extreme quantile regression neural networks}\label{s:eqrnn}

In this section we propose a new methodology that combines the extrapolation power of the GPD model with the high-dimensional predictor space capabilities and flexibility of neural networks to obtain accurate estimates for quantile functions $Q_{\bm{x}}(\tau)$ at extreme levels $\tau$. Let $\mathcal{D}=\left\{(\bm{x}_i, y_i)\right\}_{i=1}^n$ be the training dataset. Estimation of conditional extreme quantiles $\hat{Q}_{\bm{x}}(\tau)$, using~\eqref{e:gpdquant}, requires estimators for the intermediate quantile function $\hat{Q}_{\bm{x}}(\tau_0)$ with $\tau_0 < \tau$ and the GPD parameter $\sigma(\bm x)$ and $\xi(\bm x)$. It is customary to proceed in two steps.

First, we model the intermediate quantile at level $\tau_0$ using classical quantile regression methods.
We then define the conditional exceedances 
$$z_i:= y_i - \hat{Q}_{\bm{x}_i}(\tau_0), \quad i \in \mathcal{I}:= \{i=1,\ldots,n \, : \; y_i > \hat{Q}_{\bm{x}_i}(\tau_0)\}.$$
The intermediate probability $\tau_0$ should be chosen low enough to allow for stable estimation of $Q_{\bm{x}}(\tau_0)$ with classical empirical methods, but high enough for the approximation in~\eqref{GPD_cond} to be accurate, so that the exceedances $z_i$ are approximate samples of a GPD. 
However, it is not a classical tuning parameter, since different values for $\tau_0$ yield different subsets of exceedances $\mathcal{I}$. Comparison of the loss function~\eqref{e:ogpdlik} on these datasets would, therefore, not be meaningful. Instead, the threshold is, in the univariate case, usually selected in terms of stability plots and sensitivity analyses.

In the second step, we estimate the GPD parameters $\sigma(\bm x)$ and $\xi(\bm x)$ based on the set of exceedances $z_i$, $i\in \mathcal I$. 
Modelling these parameters directly in the extrapolation formula~\eqref{e:gpdquant} may lead to strong dependence between the estimates and numerical instabilities. We, therefore, rely on an orthogonal reparametrization that has a diagonal Fisher information matrix. As for the standard asymptotic GPD likelihood properties, the latter is well-defined for the GPD model, when $\xi(\bm x)>-0.5$, and the reparametrization 
$$(\sigma(\bm x),\xi(\bm x))\mapsto (\nu(\bm x),\xi(\bm x)), \qquad \nu(\bm x):=\sigma(\bm x) (\xi(\bm x)+1)$$ 
yields the desired orthogonality~\citep{orthoGPD, ExGAM}. In our experiments this reparametrization significantly improves stability and convergence in every considered setting. 

In this section we propose a flexible neural network model for the orthogonalized GPD parameters $\nu(\bm x;\mathcal{W})$ and $\xi(\bm x;\mathcal{W})$, where $\mathcal W$ denotes the collection of all model parameters. 
This can be seen as conditional density estimation with output dimension $q=2$, where the parametric family is the GPD model with parameters $\theta = (\nu, \xi)$ depending on the covariate $\bm X = \bm x$. In general, an estimate of the model parameters $\hat{\mathcal W}$ is thus obtained as a minimizer of the GPD deviance loss over the training exceedances
\begin{equation}\label{e:eqrnnoptim}%
\hat{\mathcal{W}}\in\argmin_{\mathcal{W}}\sum_{i\in\mathcal{I}}\ell_{\rm OGPD}\{z_i; \hat{\nu}(\bm{x}_i;\mathcal{W}),\hat{\xi}(\bm{x}_i;\mathcal{W})\},
\end{equation}%
where the deviance or negative log-likelihood of the GPD  in terms of the orthogonal reparametrization is 
\begin{equation}\label{e:ogpdlik}%
\ell_{\rm OGPD}(z;\nu,\xi)= \left(1+\dfrac{1}{\xi}\right) \log\left\{1+\xi\dfrac{(\xi +1)z}{\nu}\right\} + \log(\nu) - \log(\xi +1).
\end{equation}%
This yields a flexible model for the conditional tail distribution of $Y\mid \bm{X} = \bm{x}$ with which not only $\hat{Q}_{\bm{x}}(\tau)$ can be regressed but also conditional exceedance probabilities over a high threshold or conditional expected shortfalls, for example.

In the next two subsections, we discuss the details of the model for independent observations and for time series data with sequential dependence, respectively.

\subsection{Independent observations}\label{ss:eqrnnind}

We first consider the case where the training data $\mathcal{D}=\left\{(\bm{x}_i, y_i)\right\}_{i=1}^n$ is a set of independent, identically distributed observations of $(\bm X,Y)$.
The goal in this case is the estimation of the conditional quantile
$ Q_{\bm x} (\tau)$ for a predictor value $\bm X = \bm x$ at an extreme level $\tau > 0$.
For the first step of estimating the intermediate quantile function, in principle, any classical quantile regression method can be used. To avoid overfitting and obtain unbiased generalization error estimates from the training set $\mathcal{D}$, the predicted $\hat{Q}_{\bm{x}_i}(\tau_0)$, $i=1,\dots, n$, should be constructed out of training sample. This is achievable in two ways.
Using bagging methods, such as generalized random forests~\citep{GRF}, is a convenient choice since they allow for out-of-bag predictions where only a single fit on $\mathcal{D}$ is required. For other methods such as quantile regression neural networks \citep{QRNN}, out of training sample predictions can be obtained in a foldwise manner similar to cross-validation. In the sequel we assume that the intermediate quantile, and thus the exceedances, are given.

In the second step, we propose to model the orthogonalized GPD parameters $\nu(\bm x)$ and $\xi(\bm x)$ by a fully-connected feed-forward neural network with parameter vector $\mathcal W$ and deviance loss function as in~\eqref{e:eqrnnoptim}; see Section~\ref{ss:cdenn} for details.
Choices of the network architecture such as the number of neurons, the number of layers and activation functions, are hyperparameters, denoted by $\Theta$, to be selected. We provide sensible default values in our implementation, but one can also choose them in a data-driven way based on a validation set. 

The only restrictions are on the output activation functions, since ${\nu}(\bm x)$ should be strictly positive. We find the exponential function or the ${\rm SELU}$ activation~\citep{SNN} shifted above zero to be good choices.
Regarding the output activation for ${\xi}$, no strict restrictions apply and the identity would be a natural choice. However, standard likelihood regularity properties are not satisfied for the GPD model when $\xi\leq -0.5$, which very rarely occurs in practice. We observe that smoothly restricting the shape estimates, for example, between $-0.5$ and $0.7$ with the activation $x\mapsto 0.6 \tanh{(x)}+0.1$, helps to improve training stability. This avoids aberrant $\xi$ estimates in the early stages of the training and still covers almost all practical cases.
In many situations it is reasonable to assume that only the scale $\nu(\bm{x})$ varies locally but $\xi(\bm x) \equiv \xi$ is constant \citep[e.g.,][]{kin2016}. This can be achieved by restricting the network so that the shape output only depends on a bias term.

Algorithm~\ref{a:eqrnnind} summarizes our extreme quantile regression network (EQRN) for independent observations, which takes the intermediate quantiles and training data as input and outputs the extreme quantile at a desired test predictor value $\bm x\in \mathbb R^p$ and level $\tau > \tau_0$. Optionally, the conditional GPD parameters can also be obtained.

\begin{algorithm}[b]
\caption{EQRN for independent observations}\label{a:eqrnnind}
The tuning parameters $\Theta$ for the conditional GPD density estimation network $g_\mathcal{W}$ and the intermediate quantile model $\hat{Q}_\cdot(\tau_0)$ capable of out-of-sample prediction are prespecified.
The training data $\mathcal{D}=\left\{(\bm{x}_i, y_i)\right\}_{i=1}^n$ and test covariates $\bm{x}$ are observed. Let $\tau\in(\tau_0,1)$ be the desired probability level.
\begin{algorithmic}[1]
\Procedure{EQRN-Fit}{$\mathcal{D}$, $\Theta$, $\hat{Q}_\cdot(\tau_0)$}%
\State $\mathcal{I} \gets \{i=1,\ldots,n \, : \; y_i > \hat{Q}_{\bm{x}_i}(\tau_0)\}$
\State $z_i \gets y_i - \hat{Q}_{\bm{x}_i}(\tau_0) \;\; \forall i \in\mathcal{I}$
\State $\mathcal{T}, \mathcal{V} \gets$ \Call{RandomValidationSplit}{$\mathcal{I}$}
\Comment{If no validation: $\mathcal{T}=\mathcal{I}, \mathcal{V}=\emptyset$}
\State $\hat{\mathcal{W}} \gets$ \Call{InitializeNetworkWeights}{$\Theta$}
\For{$e = 1$ to maximum number of epochs $E$}
\ForAll{$\mathcal{B}\in$ \Call{GetMiniBatches}{$\mathcal{T}$}}
\State $\{(\hat{\nu}_i,\hat{\xi}_i)\}_{i\in\mathcal{B}} \gets$ $g_{\hat{\mathcal{W}}}(\bm{x}_\mathcal{B}, \hat{Q}_{\bm{x}_\mathcal{B}}(\tau_0))$ 
\State $\ell \gets \sum_{i\in\mathcal{B}}\ell_{\rm OGPD}(z_i, \hat{\nu}_i,\hat{\xi}_i)/\abs{\mathcal{B}}$
\State $\hat{\mathcal{W}} \gets$ \Call{BackPropUpdate}{$\ell$, $\hat{\mathcal{W}}$, $\bm{x}_\mathcal{B}$, $\hat{Q}_{\bm{x}_\mathcal{B}}(\tau_0)$, $\Theta$}
\EndFor
\State {\bf stop if} $\mathcal{V}\neq\emptyset$ and \Call{LossNotImproving}{$\hat{\mathcal{W}}$, $\bm{x}_\mathcal{V}$, $\hat{Q}_{\bm{x}_\mathcal{V}}(\tau_0)$, $z_\mathcal{V}$}
\EndFor
\State {\bf output} $\hat{\mathcal{W}}$ %
\EndProcedure
\vspace{2mm}
\Procedure{EQRN-Predict}{$\bm{x}$, $\tau$, $\hat{\mathcal{W}}$, $\hat{Q}_\cdot(\tau_0)$}%
\State $\{\hat{\nu}(\bm{x}),\hat{\xi}(\bm{x})\} \gets$ $g_{\hat{\mathcal{W}}}(\bm{x}, \hat{Q}_{\bm{x}}(\tau_0))$
\State $\hat{\sigma}(\bm{x}) \gets \hat{\nu}(\bm{x})/\{\hat{\xi}(\bm{x}) + 1\}$
\State compute $\hat{Q}_{\bm{x}}(\tau)$ w.r.t.\ $\hat{\sigma}(\bm{x})$, $\hat{\xi}(\bm{x})$, $\hat{Q}_{\bm{x}}(\tau_0)$, $\tau$ and $\tau_0$ using equation~\eqref{e:gpdquant}
\State {\bf output} $\hat{Q}_{\bm{x}}(\tau)$, and optionally $\{\hat{\sigma}(\bm{x}),\hat{\xi}(\bm{x})\}$
\EndProcedure
\end{algorithmic}
\end{algorithm}

The conditional GPD estimation in the second step relies on the exceedances to ensure that only information from the tail is used for extrapolation. There may, however, be residual information in the moderately extreme observations that should not be discarded. We propose to use the intermediate quantiles $\hat{Q}_{\bm{x}_i}(\tau_0)$ as an additional feature in the conditional density estimation. This feature engineering seems to consistently and significantly improve the accuracy of the final prediction $\hat{Q}_{\bm{x}}(\tau)$ on test data in our simulations. This idea is, to some degree, related to stacked learning~\citep{StackedRegr, StackedGen} and is achieved by considering the $(p+1)$-dimensional vector $(\bm{x}_i,\hat{Q}_{\bm{x}_i}(\tau_0))$, $i = 1,\dots, n,$
instead of $\bm{x}_i$ as predictors, for the network input. To avoid overfitting, it is again important that the intermediate quantile estimates are constructed out of training sample.
This new feature also improves other extreme quantile regression methods, such as the GBEX model~\citep{gbex}, as observed in our simulations in Section~\ref{s:sims}.

Classical backpropagation and optimization procedures are performed to find $\hat{\mathcal{W}}$. To avoid overfitting and local minima convergence, the validation loss can be tracked as discussed in Section~\ref{ss:cdenn}. The hyperparameters  $\Theta$ of this network include choices of optimization algorithms and regularization, for instance.
More details on the function calls in the algorithm can be found in Supplementary Material~\ref{ss:functions}.

Since the true quantile $Q_{\bm{x}}(\tau)$ is unknown in real-world data, we can not assess the performance of $\hat{Q}_{\bm{x}}(\tau)$ with metrics such as mean squared or absolute errors. As illustrated in Section~\ref{ss:qr}, the quantile loss is also unreliable due to the data scarcity at extreme quantiles. We, therefore, choose the final validation loss based on the GPD deviance to compare different choices of hyperparameters, as it is the most reliable surrogate metric.

\subsection{Sequential dependence}\label{ss:eqrnnrec}

In many applications the observations are not independent but display sequential dependence, such as in time series. In this case we denote the training data by $\mathcal{D}=\left\{(\bm{x}_t, y_t)\right\}_{t=1}^T$, which are observed sequentially from a time series $\left\{(\bm{X}_t, Y_t)\right\}_{t=1}^T$.
The goal here is different from the case of independent observations. Indeed, we would like to predict as well as possible high quantiles of the response $Y_{u}$ at some time point $u$ one step in the future based on all past information $\tilde{\bm{X}}_{u}:=\left\{(\bm{X}_t, Y_t)\right\}_{t<u}$. Therefore, the target is
\begin{equation}\label{e:condquantile_time}%
Q_{\tilde{\bm x}_{u}}(\tau) :=F^{-1}_{Y_{u} \mid  \tilde{\bm{X}}_{u} = \tilde{\bm{x}}_{u} } (\tau),
\end{equation}%
where $\tilde{\bm{x}}_{u}:=\left\{(\bm{x}_t, y_t)\right\}_{t<u}$ are observations that are not necessarily part of the training set.
In this section we propose a recurrent neural network to solve this task. Several principles are the same as in the case of independent observations, such as the choice of output activation functions. We thus focus on the differences to the independent case.

While in principle it is still possible to use any classical quantile regression method to model the intermediate conditional quantiles at level $\tau_0$, we recommend using quantile regression neural networks~\citep{QRNN,iQRNN} with a recurrent architecture. These models are specifically designed for sequential dependence and can easily adapt to varying sequence lengths of the input features $\tilde{\bm{X}}_{t}$. In our experiments, recurrent quantile regression neural networks consistently outperformed generalized random forests in the presence of sequential dependence.
Although a varying sequence length of input features is possible, we restrict this length to a fixed horizon $s \ll T$ for computational efficiency. Thus, for any time point $t$, we define its past by $\tilde{\bm{x}}_t =\{(\bm{x}_j,y_j)\}_{j=t-s}^{t-1}$. For simplicity, we denote our augmented training set by $\tilde{\mathcal{D}}=\{(\tilde{\bm{x}}_t, y_t)\}_{t=s+1}^T$.

\begin{algorithm}[t]
\caption{EQRN for sequential observations}\label{a:eqrnnrec}
The tuning parameters $\Theta$ for the recurrent conditional GPD density estimation network $\tilde{g}_\mathcal{W}$, the intermediate quantile model $\hat{Q}_\cdot(\tau_0)$ capable of out of sample prediction and horizon $s$ are prespecified.
The training data $\tilde{\mathcal{D}}=\left\{(\tilde{\bm{x}}_t, y_t)\right\}_{t=s+1}^T$ and test covariates $\tilde{\bm{x}}_u$ are observed. Let $\tau\in(\tau_0,1)$ be the desired probability level. 
\begin{algorithmic}[1]
\Procedure{EQRN-Fit}{$\tilde{\mathcal{D}}$, $\tau_0$, $s$, $\Theta$, $\hat{Q}_\cdot(\tau_0)$}%
\State $z_t \gets y_t - \hat{Q}_{\tilde{\bm{x}}_t}(\tau_0) \;\; \forall t \in\mathcal{I}:=\{t=s+1,\ldots,T \, : \; y_t > \hat{Q}_{\tilde{\bm{x}}_t}(\tau_0)\}$
\State $\mathcal{T}, \mathcal{V} \gets$ \Call{SequentialValidationSplit}{$\mathcal{I}$}
\State $\hat{\mathcal{W}} \gets$ \Call{InitializeRecurrentNetWeights}{$\Theta$}
\For{$e = 1$ to maximum number of epochs $E$}
\ForAll{$\mathcal{B}\in$ \Call{GetMiniBatches}{$\mathcal{T}$}}
\State $\{(\hat{\nu}_t,\hat{\xi}_t)\}_{t\in\mathcal{B}} \gets$ $\tilde{g}_{\hat{\mathcal{W}}}(\tilde{\bm{x}}_\mathcal{B}, \hat{Q}_{\tilde{\bm{x}}_\mathcal{B}}(\tau_0))$ 
\State $\ell \gets \sum_{t\in\mathcal{B}}\ell_{\rm OGPD}(z_t, \hat{\nu}_t,\hat{\xi}_t)/\abs{\mathcal{B}}$
\State $\hat{\mathcal{W}} \gets$ \Call{BackPropUpdate}{$\ell$, $\hat{\mathcal{W}}$, $\tilde{\bm{x}}_\mathcal{B}$, $\hat{Q}_{\tilde{\bm{x}}_\mathcal{B}}(\tau_0)$, $\Theta$}
\EndFor
\State {\bf stop if} $\mathcal{V}\neq\emptyset$ and \Call{LossNotImproving}{$\hat{\mathcal{W}}$, $\tilde{\bm{x}}_\mathcal{V}$, $\hat{Q}_{\tilde{\bm{x}}_\mathcal{V}}(\tau_0)$, $z_\mathcal{V}$}
\EndFor
\State {\bf output} $\hat{\mathcal{W}}$ %
\EndProcedure
\vspace{2mm}
\Procedure{EQRN-Predict}{$\tilde{\bm{x}}_u$, $\tau$, $\hat{\mathcal{W}}$, $\hat{Q}_\cdot(\tau_0)$}%
\State $\{\hat{\nu}(\tilde{\bm{x}}_{u}),\hat{\xi}(\tilde{\bm{x}}_{u})\} \gets$ $\tilde{g}_{\hat{\mathcal{W}}}(\tilde{\bm{x}}_{u}, \hat{Q}_{\tilde{\bm{x}}_{u}}(\tau_0))$ 
\State $\hat{\sigma}(\tilde{\bm{x}}_{u}) \gets \hat{\nu}(\tilde{\bm{x}}_{u})/\{\hat{\xi}(\tilde{\bm{x}}_{u}) + 1\}$
\State compute $\hat{Q}_{\tilde{\bm{x}}_{u}}(\tau)$ w.r.t.\ $\hat{\sigma}(\tilde{\bm{x}}_{u})$, $\hat{\xi}(\tilde{\bm{x}}_{u})$, $\hat{Q}_{\tilde{\bm{x}}_{u}}(\tau_0)$, $\tau$ and $\tau_0$ using equation~\eqref{e:gpdquant}
\State {\bf output} $\hat{Q}_{\tilde{\bm{x}}_{u}}(\tau)$, and optionally $\{\hat{\sigma}(\tilde{\bm{x}}_{u}),\hat{\xi}(\tilde{\bm{x}}_{u})\}$
\EndProcedure
\end{algorithmic}
\end{algorithm}

Algorithm~\ref{a:eqrnnrec} summarizes our EQRN for sequential data.
For the estimation of the conditional GPD parameters, the main difference compared to the independent model is the use of a recurrent architecture to capture the sequential nature of the data. If validation splitting is used during training, the split should preserve the sequential structure instead of being performed randomly.
For the use of the intermediate quantile as a feature, two approaches seem relevant. The first one is to only use $\hat{Q}_{\tilde{\bm{x}}_t}(\tau_0)$ as a separate additional input to $\tilde{\bm{x}}_{t}$ in the network. The second approach is to also use past intermediate information by considering $\{(\bm{x}_j,y_j,\hat{Q}_{\tilde{\bm{x}}_j}(\tau_0))\}_{j=t-s}^{t-1}$, instead of $\tilde{\bm{x}}_{t}$, as input features to model the GPD parameters $\nu(\tilde{\bm{x}}_t)$ and $\xi(\tilde{\bm{x}}_t)$. We prefer the second approach, as it can pass more information to the tail model.
More details on the function calls in the algorithm can be found in Supplementary Material~\ref{ss:functions}.

The training data consists of time points $\{1,\dots , T\}$, while the test data uses information from time points $\{u-s, \dots, u-1\}$ to predict at time $u$. These two intervals are typically disjoint when the model was fitted in the past and is applied for prediction in the present. The prediction model can, of course, be used to predict on the training data when $u \leq T$, but such predictions might be overly precise since $y_u$ was used in the training procedure.

\section{Simulation study}\label{s:sims}

\subsection{Setup}\label{ss:simssetup}
In this section we assess the accuracy of our EQRN model in predicting extreme conditional quantiles on simulated data and compare it to existing state-of-the-art methods. The aim is to study a simplified version of the application, which motivates the simulation setup and modelling choices. We thus focus on the case of sequentially dependent data; a simulation study for independent data can be found in Supplementary Material~\ref{ss:simsiid}.

The main competitors from the extreme value literature in terms of flexibility are the generalized additive models (EGAM)~\citep{ExGAM2} and gradient boosting for extreme quantile regression (GBEX)~\citep{gbex}, which both use conditional GPD modelling. For sequentially dependent data, we also consider the extreme quantile autoregression (EXQAR)~\citep{EXQAR} as, although assuming linear quantile dependence, the model is designed for time series. As a benchmark we consider an unconditional GPD model that ignores the covariate dependence altogether and a semiconditional GPD model that uses the covariate only for the intermediate quantile. We include results for the generalized random forests for quantile regression (GRF)~\citep{GRF}, which does not use extrapolation for high quantiles.
The training data $\mathcal{D}=\left\{(\bm{x}_t, y_t)\right\}_{t=1}^T$, with $T=\numprint{7000}$, are sequentially generated from the time series
\begin{equation}\label{e:tssim}
\begin{cases}
Y_t = \sigma_t\abs{\varepsilon_t^Y}, \quad X_t = 0.4\cdot X_{t-1} + \abs{\varepsilon_t^X}, \quad \varepsilon_t^Y, \varepsilon_t^X \sim \mathcal{N}(0,1),\\
\begin{split}
\sigma_t^2 = 1 &+ 0.1\cdot\{2Y_{t-1}^2 + Y_{t-2}^2 + Y_{t-3}^2 + Y_{t-4}^2 + Y_{t-5}^2\} +\\
&+ 0.1\cdot\{3X_{t-1}^2 + 2X_{t-2}^2 + X_{t-3}^2 + X_{t-4}^2 + X_{t-5}^2\}.
\end{split}
\end{cases}
\end{equation}
Figure~\ref{f:tsdata} in Supplementary Material~\ref{ss:simsseqsuppl} shows part of the simulated data.
To have a fair comparison, all methods use the same covariate vectors $\tilde{\bm{x}}_t =\{(\bm{x}_j,y_j)\}_{j=t-s}^{t-1}$ with $s=10$.
This model admits a GPD approximation as in~\eqref{GPD_cond} since the conditional distribution $Y_t \mid \tilde{\bm{X}}_t =\tilde{\bm{x}}_t $ is a folded normal distribution, and its tail can, therefore, be approximated by a GPD with shape parameter $\xi(\tilde{\bm{x}}_t) = 0$.
For the methods that use covariate-dependent intermediate quantiles, we use the same estimates $\hat{Q}_{\tilde{\bm{x}}_t}(\tau_0)$ with $\tau_0 = 80\%$ from a recurrent quantile regression neural network (QRN); for a sensitivity analysis of the choice of $\tau_0$, see Supplementary Material~\ref{ss:simsseqsuppl}. The best QRN architecture and hyperparameters are chosen based on validation quantile loss. For the methods that use covariate-dependent GPD parameters, we also use $\hat{Q}_{\tilde{\bm{x}}_t}(\tau_0)$ as additional covariate. Although designed for univariate time series, we adapted EXQAR to accept several covariate sequences. Those two choices significantly improve the competitors' performances.

For the EQRN model, $\numprint{2000}$ observations are kept for validation tracking, and thus only the remaining $\numprint{5000}$ are effectively used for weight training.
The best choices for EQRN hyperparameters are made based on validation loss by performing a grid search over a set of possible values and network architectures. 
All other models are fitted on the whole training dataset, as they do not use validation loss tracking. The best set of hyperparameters for GBEX (tree depths, learning rate and number of trees) are chosen using cross-validation, and the ground truth for whether the shape is constant is given to EGAM. For EXQAR we use $\delta_{2n}=n^{-0.9}$, as recommended by the authors, but set $\delta_{1n}=1-\tau_0$, thus increasing the number of quantile pseudo-observations used for inference and allowing better comparison with the other methods.
The predictions of all models are evaluated by their root mean squared error (RMSE) compared to the true conditional quantiles on a newly generated test dataset that follows the same distribution as the training data in~\eqref{e:tssim}.

\subsection{Results}\label{ss:simsts}

For estimation of the conditional GPD parameters with our recurrent EQRN model, we consider both LSTM and GRU architectures with one to three recurrent layers and hidden dimensions between $32$ and $256$. As the networks are not too deep, $L_2$ penalty was chosen over dropout for regularization during training, with possible penalty $\lambda\in\{0,10^{-6},10^{-5},10^{-4},10^{-3}\}$. Both constant and covariate-dependent shape parameter outputs are considered. The model with minimum validation loss is a single LSTM layer with hidden dimension $128$, followed by the usual fully connected output layer, with constant shape and $\lambda=10^{-4}$. As a comparison we also retain predictions for the best unpenalized network with $\lambda = 0$ and fixed shape, which has two LSTM layers of hidden size $128$.

\begin{figure}[t]
\centering
\includegraphics[width=\textwidth]{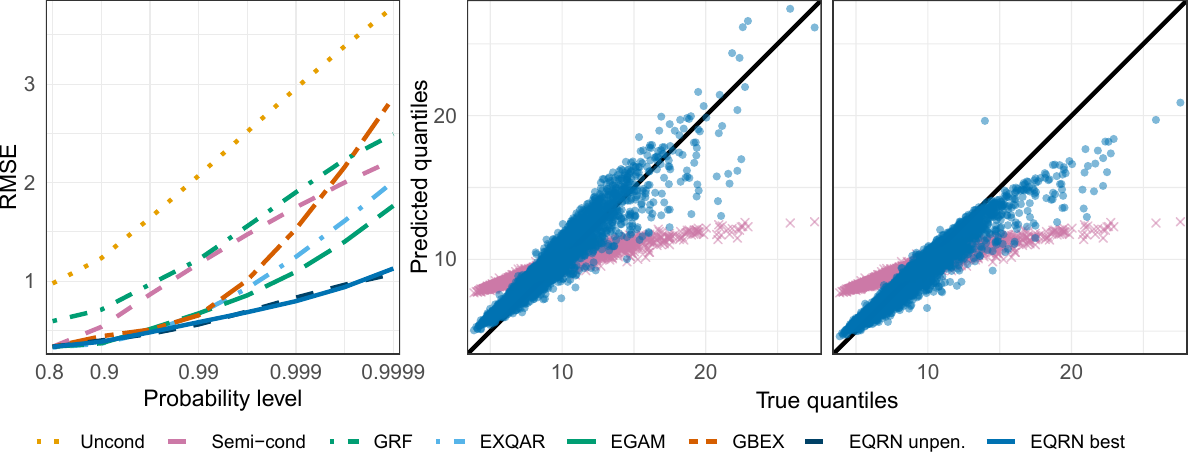}%
\caption{Left: Root mean squared error between predicted and true conditional quantiles at different probability levels $\tau$ (log-scale), for the selected EQRN models and the improved competitors. Centre-right: True vs.\ predicted quantiles at probability level $\tau=99.95\%$ for the best unpenalized (middle) and penalized (right) EQRN models (dots), compared to the semiconditional estimates (crosses).}%
\label{f:tscompall}
\end{figure}

The left panel of Figure~\ref{f:tscompall} shows the RMSE of best penalized and unpenalized EQRN models, compared to the improved competitors, as a function of the quantile level $\tau$. We observe that, for the lowest level $\tau =\tau_0 = 80\%$, all structured GPD models have the same performance since they use the same intermediate quantiles. GRF and the unconditional model have already higher errors since they are not able to capture the sequential dependence at the intermediate level sufficiently.
For growing quantile levels $\tau$, the errors of the covariate-dependent GPD models start to diverge. This is due to the differences in modelling flexibility in terms of the GPD parameters of each method. We observe a similar behaviour for EXQAR, as its linear quantile dependence is not flexible enough. Our EQRN method based on recurrent neural networks seems to be best at modelling sequential tail dependence.

Figure~\ref{f:tscompall} also shows the predicted quantiles $\hat Q_{\tilde{\bm{x}}_u}(\tau)$ on the test data compared to the true $Q_{\tilde{\bm{x}}_u}(\tau)$ for a fixed $\tau = 99.95\%$ for the best penalized and unpenalized EQRN models. In general, both models seem to perform well in predicting the high conditional quantiles. 
The weight penalty seems to mainly affect the larger quantile predictions. Compared to the unpenalized model, we observe that the reduction in the variance of the predictions comes at the cost of a bias for larger quantile values. This bias-variance trade-off is typical with penalization.
The poor performance of the semiconditional estimates highlights the added value of covariate dependence in the GPD parameters.

As discussed in Section~\ref{s:eqrnn}, the choice of the intermediate level $\tau_0$ generally has an impact on the prediction accuracy. In our covariate-dependent setting, where the model for the conditional GPD is a flexible regression model, this choice seems to have less importance. Indeed, even for a fairly low value of $\tau_0$, the flexibility of the neural network model seems to be able to absorb some of the approximation bias; see 
Supplementary Material~\ref{ss:simsseqsuppl} for details.

Additional results on the quantile R squared coefficient and bias-variance decomposition of the RMSEs presented in Figure~\ref{f:tscompall} are also discussed in Supplementary Material~\ref{ss:simsseqsuppl}.

\section{Application}\label{s:appl}

\subsection{Motivation}\label{app:mot}

Flood risk is a major natural hazard in Europe, which causes huge economic damage and endangers human lives. 
There is a longstanding interest in statistical methods of extreme value theory for hydrology \citep[e.g.,][]{kat2002, kee2009, AsadiDanube, eng2018}, and national agencies commonly use them to assess the long-term risk of flooding in cities, at power plants, and other key locations. Return levels with long return periods can be estimated using the GEV distribution for annual maxima or the GPD model for daily threshold exceedances. The output then guides effective long-term flood management measures.

\begin{figure}[t]
\centering
\includegraphics[width=0.85\textwidth]{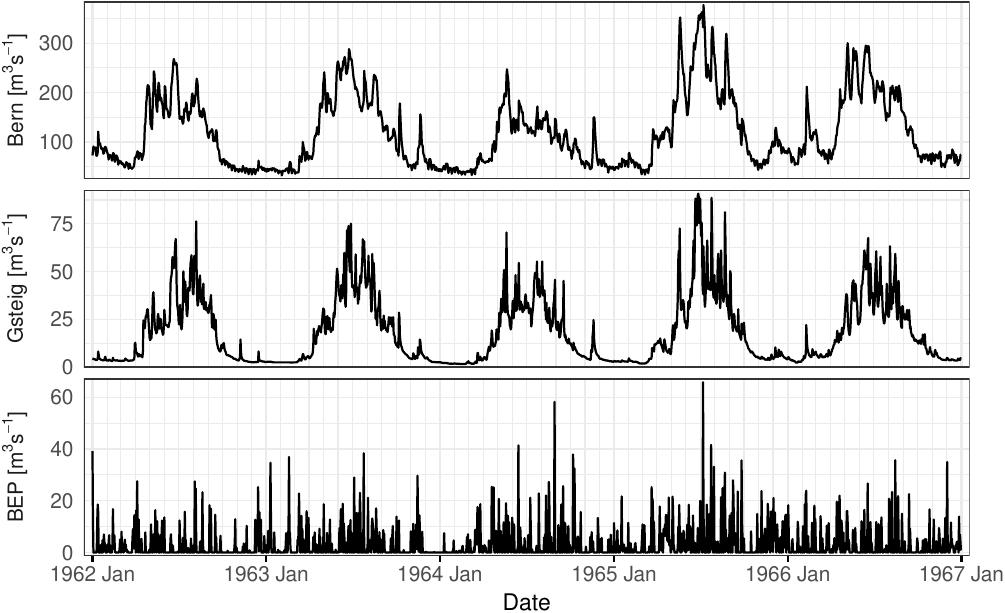}%
\caption{Daily average discharge observations at Bern--Sch\"onau (62) and at the upstream station at Gsteig (42), and daily precipitation at the closest meteorological station to Bern (BEP), over five years; see Figure~\ref{f:chstations} for geographical locations of the gauging stations.}
\label{f:chdata}
\end{figure}

An example of an important location in Switzerland is the gauging station in Bern on the Aare river, which is shown within its water catchment in Figure~\ref{f:chstations}. The Swiss Federal Office for the Environment (FOEN) monitors the Aare, and we use daily average discharges (in $\text{m}^3\text{s}^{-1}$) in Bern and another upstream station together with recordings of daily precipitation (in $\text{mm}$) at six locations in the Bern catchment; see Figure~\ref{f:chstations} for details. All time series are available in the period from 1930--2014 and can be obtained from the FOEN\footnote{\href{https://www.hydrodaten.admin.ch/}{\texttt{https://www.hydrodaten.admin.ch/}}.} (for discharges) and MeteoSwiss\footnote{\href{https://gate.meteoswiss.ch/idaweb}{\texttt{https://gate.meteoswiss.ch/idaweb}}.} (for precipitation). Figure~\ref{f:chdata} shows an excerpt for the two river stations and one precipitation gauge.

\begin{figure}[t]
\centering
\includegraphics[width=\textwidth]{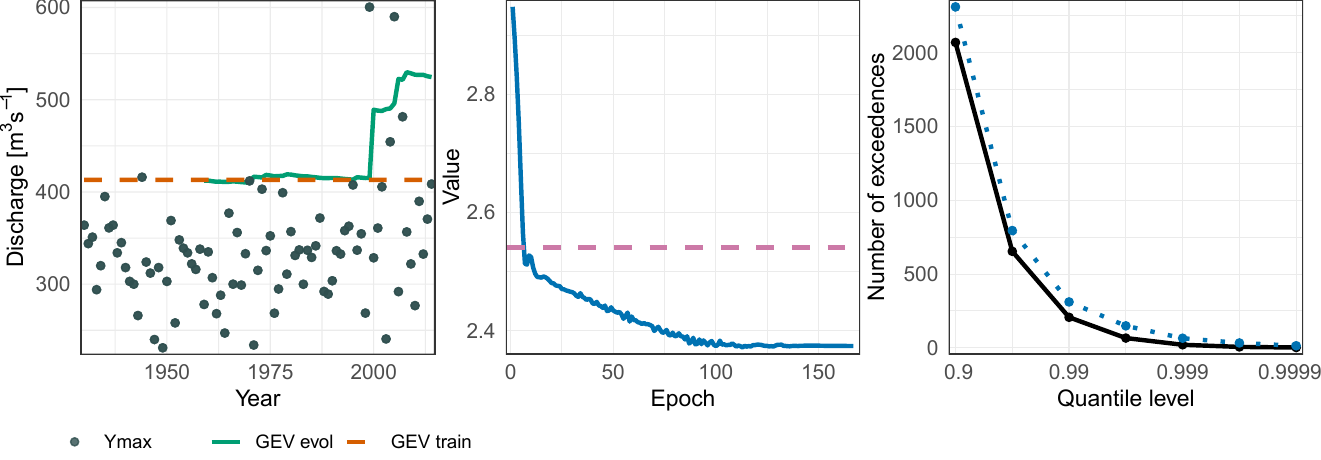}%
\caption{Left: Annual maxima of daily average discharges (points) at Bern--Sch\"onau (62) together with the unconditional 100-year return level based on GEV fitted on the training data of 1930--1958 (dashed line), and the evolution of the same return level (solid line) using data from 1930--$y$, where $y \in \{1959, \dots, 2014\}$ denotes end of the period. Middle: Evolution of the validation loss (solid line) of the selected EQRN network for the river discharge data as a function of the training epoch; the dashed line shows the validation loss of the semiconditional model. Right: Number of observations exceeding the EQRN quantile predictions on the test set (dotted line) compared to the expected number of exceedances (solid line) for different probability levels (log-scale).}%
\label{f:chnonstaticalibr}
\end{figure}

To illustrate possible drawbacks of a classical extreme value analysis, the left panel of Figure~\ref{f:chnonstaticalibr} shows the annual maxima of river discharges at the Bernese station on the Aare. The dashed line is the estimated 100-year return level based on the GEV approximation using the training period from 1930--1958. One can see that starting from the year 1999 there are several exceedances over this return level, somewhat contradicting the fact that it should only be exceeded on average once in 100 years. The solid line is the same return level based on data from 1930--$y$, where $y \in \{1959, \dots, 2014\}$ denotes the end of the training period. While the predictions are fairly stable until 1999, an extreme value analysis performed after that year would yield much higher values for the 100-year return level. Conversely, historical estimates before such a break-point would severely underestimate the flood risk. 
In general, distributional shifts can be due to climate change, changes in the river system, or other structural breaks in factors influencing discharge at this location.
For the Bernese station, the FOEN indeed reports a significant break-point in extreme discharges in the nineties but acknowledges that a clear cause can not be identified\footnote{See flood report of the FOEN at \href{https://gate.meteoswiss.ch/idaweb}{\texttt{https://www.hydrodaten.admin.ch/en/2135.html}}.}. One factor may be a multidecadal variability of flood occurrence, as described in \cite{sch2010}.

The aim of our methodology is complementary to classical extreme value analysis and addresses this issue with static return levels. 
We apply our EQRN model to estimate one-day-ahead extreme quantiles of the river discharge conditionally on previous observations of discharge and precipitation in the catchment. This allows the forecasting of flood risk even in nonstationary systems, such as a changing climate. The strength of our approach lies in the ability to exploit information from multiple covariates and capture the complex time dependence. Even in situations where the causes for structural changes are unknown, our method implicitly accounts for them through their effects on the covariates.
The output of the model can help practitioners and authorities to manage flooding more effectively and help to minimize their disastrous impacts by early warning systems.

To illustrate our methodology and show its effectiveness in comparison to classical forecasting approaches, we consider in more detail the flood in August 2005 in Switzerland. 
At the Bernese gauging station, it was the largest event since the beginning of the recordings, and it caused severe economic damage across large parts of the country and the loss of several lives.

\subsection{Model specification}

The whole dataset consists of 31,046 daily observations $(\bm x_t, y_t)$ between 1930--2014. The response $y_t$ is the daily average discharge at the Bernese gauging station on the Aare, and the covariates $\bm x_t \in \mathbb R^p$, $p=7$, consist of discharge at another upstream station and daily precipitation measurements from six locations in the same catchment; see Figures~\ref{f:chstations} and~\ref{f:chdata}.
The discharges show significant seasonality, both in trend and variance, with the largest extremes only appearing in the summer. We do not reduce artificially the nonstationarity via classical approaches from times series analysis \citep[e.g.,][]{STL}, as we believe the seasonality and other trends are captured through the covariates.

We split the data into training and test sets. The first $T = 10,349$ observations in the period between 1930--1958 are used to train the models, whereof the first three-quarters serve the parameter estimation, and the remaining quarter is a validation set to determine hyperparameters (sets $\mathcal T$ and $\mathcal V$ in Algorithm~\ref{a:eqrnnrec}, respectively). The test set contains \numprint{20697} observations from 1958--2014, which is used for neither fitting nor selection of parameters but only to evaluate the model performance on an independent time period.
We choose this rather small proportion of training data to study the ability of the model to adapt to possible nonstationarity over time without refitting. In particular, the model weights are not updated with any information from data after 1958, even for forecasts in the study of the 2005 flood of interest. A large test set is also required to evaluate extreme properties of the data distribution. 
As augmented covariates at time $t$ for the recurrent neural network models, we use the $s=10 $ preceding days and set $\tilde{\bm{x}}_t =\{(\bm{x}_j,y_j)\}_{j=t-s}^{t-1}$. The augmented training set is then $\tilde{\mathcal{D}}=\{(\tilde{\bm{x}}_t, y_t)\}_{t=s+1}^T$.
Only observations of the $p+1=8$ variables during the preceding 10 days are used to predict one day ahead for a new test time point.

As discussed in Section~\ref{s:eqrnn}, we perform two steps for the estimation of the conditional tail model. First, we fit an intermediate quantile regression model to estimate $Q_{\tilde{\bm x}_t}(\tau_0)$. As in the simulations with sequential dependence, we choose a recurrent QRN for this purpose and set $\tau_0 = 0.8$. For the second step, we include the intermediate quantile estimates $\hat Q_{\tilde{\bm x}_j}(\tau_0)$ during the same time horizon $j=t-s, \dots, t-1$ as additional covariates, and, slightly abusing notation,
we denote $\tilde{\bm x}_t$ as the new covariate vector; see Section~\ref{ss:eqrnnrec}.
A recurrent EQRN is then fitted to the exceedances for estimation of the conditional GPD parameters $\sigma(\tilde{\bm x}_t)$ and $\xi(\tilde{\bm x}_t)$; see Algorithm~\ref{a:eqrnnrec}.
Similarly to the study in Section~\ref{ss:simsts}, a grid search is performed on the training data to select the best hyperparameters and architectures for both recurrent neural network models based on validation losses.

The final model chosen to regress the intermediate quantiles is a QRN with two LSTM layers of dimension $256$, followed by the usual fully connected layer, and $L_2$ weight penalty with parameter $\lambda=10^{-6}$. The chosen EQRN model has two LSTM layers of dimension $16$, followed by a fully connected layer, and $L_2$ weight penalty with parameter $\lambda=10^{-6}$.

\subsection{Results}\label{app:results}
The middle panel of Figure~\ref{f:chnonstaticalibr} shows the validation loss (solid line) of the selected EQRN model as a function of the training epoch. It can be seen that already after a few epochs, the method has a lower loss than the simple semiconditional model with constant GPD parameters $\sigma$ and $\xi$ (dashed line). This shows that the GPD distribution varies with the predictor values $\tilde{\bm x}_t$ and that a flexible model is beneficial.

The main output of the EQRN model are the extreme quantile estimates $\hat Q_{\tilde{\bm x}_u}(\tau)$ for a level $\tau$ and a time point $u$ of interest, conditionally on the past covariates $\tilde{\bm x}_u$. These one-day-ahead risk forecasts are shown as a function of time on the test set in the top panel of Figure~\ref{f:chpreds2005}. We observe that the model is able to extrapolate beyond the range of the data since the event shown in the plot is unprecedented and the predictions still anticipate the first exceedance of the unconditional 100-year return level $Q^{100}$.
 
An unconditional $\tau$-quantile is defined as the value that is exceeded by a proportion of $1-\tau$ of the data. An analogous property holds for conditional quantiles in data with sequential dependence, which yields a natural model assessment tool.
On the population level, if $(\tilde{\bm X}_t, Y_t)_{t=1}^T$ is the random time series with augmented covariate vectors, then the expected number of exceedances over the true conditional $\tau$-quantiles $Q_{\tilde{\bm X}_t}(\tau)$ is 
$$\mathbb E \sum_{t=1}^T 1\{ Y_t > Q_{\tilde{\bm X}_t}(\tau)\} = (1-\tau) T.$$
Consequently, plugging in the data $(\tilde{\bm x}_t, y_t)_{t=1}^T$ and estimates $\hat Q_{\tilde{\bm x}_t}(\tau)$ from a quantile regression method, the equation should approximately hold if the model is well-calibrated. Such a model assessment plot is shown in the right-hand panel of Figure~\ref{f:chnonstaticalibr} for our EQRN fit as a function of the quantile level~$\tau$. We observe that the model is fairly well-calibrated, with a slight bias toward more exceedances than expected. 

An additional output are the corresponding GPD parameters $\sigma(\tilde{\bm x}_u)$ and $\xi(\tilde{\bm x}_u)$, which together with the intermediate quantile $\hat Q_{\tilde{\bm x}_u}(\tau_0)$ specify the whole tail of the distribution of $Y_u \mid \tilde{\bm X}_u = \tilde{\bm x}_u$ according to~\eqref{GPD_cond}. For a given threshold level of interest $Q$, we can plot the flood risk for the next day as the one-day-ahead forecast of the exceedance probability over $Q$, that is, an estimate of the function $ u \mapsto \mathbb P( Y_u > Q \mid \tilde{\bm X}_u = \tilde{\bm x}_u)$.
The bottom panel of Figure~\ref{f:chpreds2005} shows the EQRN-based estimate of this function on the test set as a ratio to the unconditional $\mathbb P( Y_u > Q )$, where the threshold $Q$ is chosen as the static 100-year return level $Q^{100}$ based on the GEV distribution fitted on the training set; this threshold is relevant since it is often used to determine the height of dams for flood management. It results in a daily measure of how likely the exceedance on the next day is compared to what was expected unconditionally. Times with large predicted probability ratios are apparently times of imminent danger that can be used as triggers for early warning systems or additional flood management measures.

\begin{figure}[t]%
\centering
\includegraphics[width=0.9\textwidth]{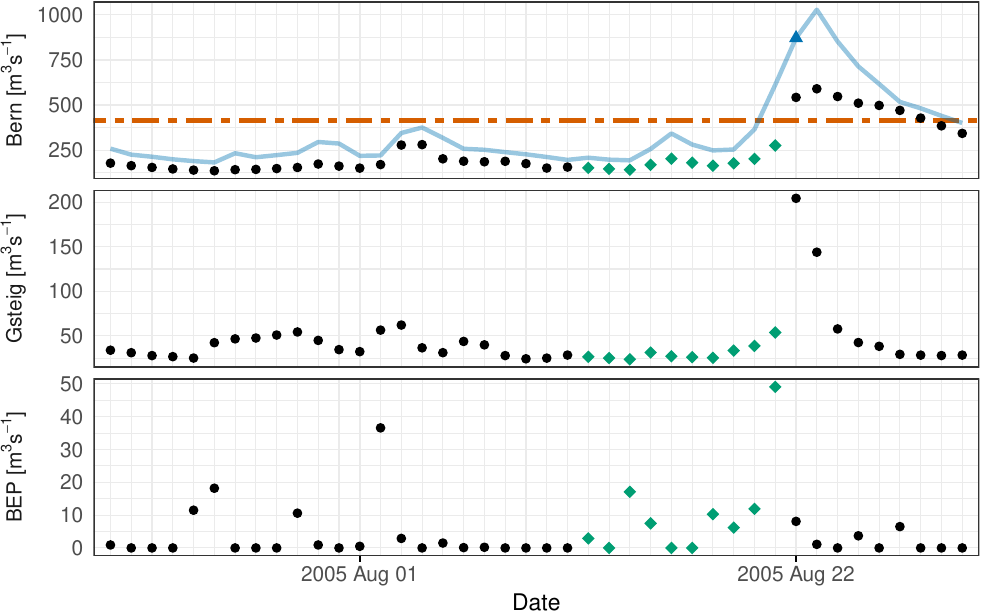}%
\caption{Discharge at Bern--Sch\"onau (62) and Gsteig (42) and precipitation at BEP during the period of the 2005 flood, where diamonds indicate covariates used for prediction of the 100-year conditional quantile (triangle) on August 22, 2005; other precipitation stations are not shown. The top panel also shows the unconditional 100-year return level (dashed line) fitted on the training data and predictions on other days (solid line).}
\label{f:chfeatspred}
\end{figure}

As an example, one may issue a warning when the forecasted conditional probability of exceeding $Q^{100}$ is, say, a hundred times larger than the baseline unconditional probability $\mathbb P( Y_u > Q^{100})$.
In the test data, there are four time clusters, which typically last several days, when $Q^{100}$ is exceeded; see Figure~\ref{f:chfeatspred} for one of these events. Applying this early warning system, in all four of these cases a timely warning would have been issued on the days preceding the first exceedance of the cluster.
Such a decision rule would lead to an average of only $1.3$ warnings for clusters of exceedances per year on the test set and is, therefore, not overly conservative. 
As the model does not need refitting on the test set, the daily forecast and possible warnings are obtained in less than one second of computation time, even on a CPU-only laptop computer\footnote{Intel Core i5-8265U 1.6 GHz processor with four cores, and eight gigabytes of RAM memory.}. The training time for the selected GPD network took less than 15 minutes.

We consider the period of the 2005 flood in Bern in more detail. Figure~\ref{f:chfeatspred} shows the two discharge time series and precipitation at the closest meteorological station before and after the event. On the evening of August 21, 2005, the day preceding the first exceedance of the 100-year GEV return level (horizontal dashed line), the prediction of our EQRN already indicated a sudden increase in the probability of this exceedance; see bottom panel of Figure~\ref{f:chpreds2005}. Equivalently, the blue point in Figure~\ref{f:chfeatspred} shows the increased value of the conditional 100-year quantile predicted by the model. The diamonds mark the observations of the previous 10 days that were used for this prediction.  

In this case the high precipitation values on August 21 and the preceding days seem to have driven this prediction, possibly together with high values of the rivers. It is interesting to note that a similar situation on August 2 has not resulted in a ``flood warning'' since the forecasted exceedance probability and return level are not exceptionally high---as a matter of fact, there was no exceedance on the next day. This means that the predictions are driven by a complex combination of the risk factors $\tilde{\bm X}_t$ in space and time that are well-captured by the recurrent architecture of the EQRN.

At the time the FOEN used an adapted version of the hydrological model HBV~\citep{hydroHBV97} for forecasting river discharges based on several inputs such as precipitation forecasts. However, the forecasts prior to this event underestimated the flood risk and resulted in too-late warnings.
Physical models for discharges and precipitation do not use explicit extrapolation in the extreme tails and often have poor performance in the largest predictions. In fact, during the 2005 flood, a main reason for the late warnings was that forecasters did not trust the predicted precipitation amounts during this extreme scenario.
In the aftermath of this flood, the FOEN, therefore, published a detailed analysis of the internal forecasting procedures~\citep{BAFU2005floodT1}.
The exact forecasts from that time are not available, but the above discussion of the results shows that our statistical methodology is a competitive alternative to physical models for the forecasting of flood risk. We also note that our model uses much less information, as it relies only on observed discharge and precipitation and does not require forecasts of atmospheric variables. 

In Section~\ref{ss:applsuppl} of the Supplementary Material, we compare and discuss the forecasts of our EQRN method with those of some of the competitors. Figures~\ref{f:CHsemicond}--\ref{f:CHgbex} show that all methods seem to capture at least some of the temporal structure based on the past covariates. Except for the GBEX method, the forecasts of the competitors suffer, however, from a low sensitivity to changes in the conditional tail or from a too erratic behaviour as a function of time. Overall, the recurrent structure of our EQRN method, therefore, seems to be the best suited model for this kind of sequentially dependent data. 
Since in real-world applications the true quantiles are unknown, direct computation of the prediction error is difficult, and model assessment plots as in the right-hand panel of Figure~\ref{f:chnonstaticalibr} are crucial.
This highlights the importance of simulation studies to evaluate and compare quantitatively the accuracy of different methods. In particular, in situations with temporal dependence, our
EQRN method clearly outperforms the competitors (e.g., Figure~\ref{f:tscompall}). This is another indicator to trust the EQRN forecasts in applications.

\section{Conclusion}\label{s:conclu}

Our EQRN model combines extrapolation results from extreme value theory with the prediction power of neural networks. It provides a flexible and versatile method for extreme quantile regression that is capable of prediction beyond the range of the data in the presence of a large number of covariates. 
Customised network architectures can be used in our open-source ``$\mathtt{EQRN}$'' R package, allowing for tailor-made models capturing all types of potential dependencies between covariates and between observations. %

The main focus in this paper was the case of sequential dependence to develop a tool for risk forecasting in time series that can be used for effective early warning systems in flood management. 
Our model already performs well in issuing sparse warnings for the days with increased risk of flooding, as illustrated in the case study of the Aare catchment in Switzerland. 
A further improvement could be attained by using additional covariates as input of the model.
This could include observations of variables that are typically used in hydrological models, such as soil moisture, or forecasts of atmospheric variables such as precipitation and temperature.

Many other applications seem pertinent. Even in the case of independent data, our simulation study in Supplementary Material~\ref{ss:simsiid} shows that neural networks outperform tree-based methods, such as ERF and GBEX, if the quantile function is more complex. Applications are financial risk assessment in insurance companies or banks.
For spatial data, images, or graphs, convolutional or graph neural networks~\citep{DLLeCun,GNN1,GNN2} are known to perform extremely well in capturing neighbourhood structures. 
Our EQRN method can, therefore, be applied to quantify the risk of climate extremes where the predictor space contains spatiotemporal observations of meteorological variables \citep[e.g.,][]{bou2022}. 
Transformer architectures~\citep{Transformers1} can also perform well for spatiotemporal or more complex dependencies.

The price for the high flexibility of machine learning methods, which focus on prediction accuracy, is limited statistical interpretability. However, feature-importance identification methods are becoming increasingly popular for interpreting neural network predictions~\citep{NNinterpr}. There is also active research on the construction of prediction intervals for black-box methods, for instance, through conformal inference~\citep{ConformalPred,ConformalizedQR}. How such techniques can be adapted to assess uncertainty for extreme quantile regression is an interesting future research question.

\section*{Declarations}
\paragraph*{Acknowledgements}
The authors would like to thank Daniel Viviroli for his valuable insights as well as the editorial team of the Annals of Applied Statistics and the anonymous reviewers for their helpful comments.

\paragraph*{Funding}
Both authors were supported by the Swiss National Science Foundation Eccellenza Grant 186858.

\paragraph*{Published article}
This document is the peer-reviewed ``Author's Accepted Manuscript'' of an article published in the Annals of Applied Statistics~\citep{Pasche2024}, with the DOI \href{https://doi.org/10.1214/24-AOAS1907}{\texttt{https://doi.org/10.1214/24-AOAS1907}}. 
When citing this work, please refer to the published version. 

\section*{Supplementary material}
\paragraph*{Supplementary results}
The Supplementary Material appended to this document contains additional information on Algorithms~\ref{a:eqrnnind} and~\ref{a:eqrnnrec}, the simulation study on independent data, additional results for the simulation study on dependent data, an analysis of the EQRN sensitivity to the intermediate probability level and competitor approaches to the application.
\paragraph*{Reproducibility and R package}
An open-source ``\texttt{EQRN}'' R package implementation of the proposed methodology is available on \href{https://github.com/opasche/EQRN}{\texttt{https://github.com/opasche/EQRN}}. 
The code and data with detailed instructions to reproduce the results presented in this paper, and more, are available on \href{https://github.com/opasche/EQRN_Results}{\texttt{https://github.com/opasche/EQRN\_Results}}.

\bibliographystyle{abbrvnat-namefirst}%
\begin{small} %
\bibliography{OCP_bibliography}
\end{small}

\newpage

\begin{appendix}

\begin{center}
{\large\bf SUPPLEMENTARY MATERIAL TO \\
\vspace{12pt}
``Neural Networks for Extreme Quantile Regression with an Application to Forecasting of Flood Risk''}
\end{center}

\setcounter{section}{0}
\setcounter{subsection}{0}
\setcounter{equation}{0}
\setcounter{figure}{0}
\setcounter{table}{0}
\renewcommand{\thesection}{S.\arabic{section}}
\renewcommand{\thesubsection}{S.\arabic{section}.\arabic{subsection}}
\renewcommand{\theequation}{S.\arabic{equation}}
\renewcommand{\thefigure}{S.\arabic{figure}}
\renewcommand{\thetable}{S.\arabic{table}}

\renewcommand{\theHsection}{S.\thesection}
\renewcommand{\theHsubsection}{S.\thesubsection}
\renewcommand{\theHequation}{S.\theequation}
\renewcommand{\theHfigure}{S.\thefigure}
\renewcommand{\theHtable}{S.\thetable}

\section{Additional LSTM illustration}\label{ss:illus}
Figure~\ref{f:MLLSTMdiagr} shows a schematic representation of a multilayer LSTM network.

\section{Details on Algorithms~\ref{a:eqrnnind} and \ref{a:eqrnnrec}}\label{ss:functions}

Algorithms~\ref{a:eqrnnind} and \ref{a:eqrnnrec} contain some abbreviated function calls. We give some details here:
\begin{enumerate}
\item[] \Call{RandomValidationSplit}{$\mathcal I$}: For independent data, splits the index set $\mathcal I$ randomly into training set $\mathcal T$ and validation set $\mathcal V$ with prespecified proportions.
\item[] \Call{SequentialValidationSplit}{$\mathcal I$}: For sequential data, splits the index set $\mathcal I$ sequentially into training set $\mathcal T$ and validation set $\mathcal V$ with prespecified proportions, such that all observations in $\mathcal T$ are before $\mathcal V$ in time.
\item[] \Call{InitializeNetworkWeights}{$\Theta$} / \Call{InitializeRecurrentNetWeights}{$\Theta$}: Initializes the weights of the GPD (recurrent) neural network randomly. The number of weights is determined by~$\Theta$. 
\item[] \Call{GetMiniBatches}{$\mathcal T$}: Splits the training set $\mathcal T$ into mini-batches for stochastic gradient descent. 
\item[] \Call{BackPropUpdate}{$\ell$, $\hat{\mathcal{W}}$, $\bm{x}_\mathcal{B}$, $\hat{Q}_{\bm{x}_\mathcal{B}}(\tau_0)$, $\Theta$}: Updates the parameter vector by a gradient step computed by backpropagation; may involve regularization methods such as $L_2$-penalty or dropout, specified in the hyperparameters $\Theta$.
\item[] \Call{LossNotImproving}{$\hat{\mathcal{W}}$, $\bm{x}_\mathcal{V}$, $\hat{Q}_{\bm{x}_\mathcal{V}}(\tau_0)$, $z_\mathcal{V}$}: If validation loss is tracked, then also a stopping criterion (e.g., for early stopping) is specified, and this function returns \texttt{TRUE} if this criterion is attained, indicating that the validation loss is not improving any more.
\end{enumerate}

\begin{figure}[t]
\centering
\begin{tikzpicture}[shorten >=1pt,->]
\definecolor{mcolr}{RGB}{213,94,0}
\definecolor{mcolb}{RGB}{0,114,178}
\definecolor{mcolg}{RGB}{0,158,115}
\definecolor{mcoly}{RGB}{230,159,0}

\def\stepsep{60pt}
\def\elabsep{9.375pt}
\def\lstmsep{90pt}
\tikzstyle{annot} = [text width=4em, text centered]
\tikzstyle{layer}=[circle, minimum size=25.0pt, fill=white, line width=1.25pt, draw=black, inner sep=1pt]
\tikzstyle{data}=[layer, fill=mcolb!75]
\tikzstyle{output}=[layer, fill=mcolg!75]
\tikzstyle{loss}=[regular polygon, regular polygon sides=4, minimum size=35.0pt, fill=mcolr!70, line width=1.25pt, draw=black]
\tikzstyle{fctarr}=[line width=1.25, ->, color=black]
\tikzstyle{usearr}=[line width=1.25, ->, dashed, color=black]
\tikzstyle{dot}=[mark size=1pt, color=black]
\tikzstyle{lstm}=[regular polygon, regular polygon sides=4, minimum size=37.0pt, fill=mcoly!20, line width=1.25pt, draw=black, inner sep=-2pt]

\node (xt) [data] at (0pt, 0pt) {$\mathbf{x}_{t-1}$};
\node (xt1) [data] at (-1*\lstmsep, 0pt) {$\mathbf{x}_{t-2}$};
\node (xt2) [data] at (-2*\lstmsep, 0pt) {$\mathbf{x}_{t-3}$};
\node (xts) [data] at (-3*\lstmsep, 0pt) {$\mathbf{x}_{t-s}$};

\node (lst) [lstm] at (0pt, \stepsep) {LSTM$_1$};
\node (lst1) [lstm] at (-1*\lstmsep, \stepsep) {LSTM$_1$};
\node (lst2) [lstm] at (-2*\lstmsep, \stepsep) {LSTM$_1$};
\node (lsts) [lstm] at (-3*\lstmsep, \stepsep) {LSTM$_1$};
\node (ht) [layer] at (0.5*\lstmsep, \stepsep+5mm) {$\mathbf{h}^{(1)}_{t-1}$};
\node (ht1) [layer] at (-0.5*\lstmsep, \stepsep+5mm) {$\mathbf{h}^{(1)}_{t-2}$};
\node (ht2) [layer] at (-1.5*\lstmsep, \stepsep+5mm) {$\mathbf{h}^{(1)}_{t-3}$};
\node (ct) [layer] at (0.5*\lstmsep, \stepsep-5mm) {$\mathbf{c}^{(1)}_{t-1}$};
\node (ct1) [layer] at (-0.5*\lstmsep, \stepsep-5mm) {$\mathbf{c}^{(1)}_{t-2}$};
\node (ct2) [layer] at (-1.5*\lstmsep, \stepsep-5mm) {$\mathbf{c}^{(1)}_{t-3}$};

\node (lspt) [lstm] at (0pt, 2*\stepsep) {LSTM$_L$};
\node (lspt1) [lstm] at (-1*\lstmsep, 2*\stepsep) {LSTM$_L$};
\node (lspt2) [lstm] at (-2*\lstmsep, 2*\stepsep) {LSTM$_L$};
\node (lspts) [lstm] at (-3*\lstmsep, 2*\stepsep) {LSTM$_L$};
\node (hpt) [layer] at (0.5*\lstmsep, 2*\stepsep+5mm) {$\mathbf{h}^{(L)}_{t-1}$};
\node (hpt1) [layer] at (-0.5*\lstmsep, 2*\stepsep+5mm) {$\mathbf{h}^{(L)}_{t-2}$};
\node (hpt2) [layer] at (-1.5*\lstmsep, 2*\stepsep+5mm) {$\mathbf{h}^{(L)}_{t-3}$};
\node (cpt) [layer] at (0.5*\lstmsep, 2*\stepsep-5mm) {$\mathbf{c}^{(L)}_{t-1}$};
\node (cpt1) [layer] at (-0.5*\lstmsep, 2*\stepsep-5mm) {$\mathbf{c}^{(L)}_{t-2}$};
\node (cpt2) [layer] at (-1.5*\lstmsep, 2*\stepsep-5mm) {$\mathbf{c}^{(L)}_{t-3}$};

\node (out) [output] at (0.5*\lstmsep, 3*\stepsep) {$\tilde{g}_\mathcal{W}(\tilde{\mathbf{x}}_{t})$};
\node (y) [data] at (0.5*\lstmsep-2*\stepsep, 3*\stepsep) {$y_{t}$};
\node (L) [loss] at (0.5*\lstmsep-1*\stepsep, 3*\stepsep) {$\ell$};

\draw [fctarr] (xt) to  (lst);
\draw [fctarr] (xt1) to  (lst1);
\draw [fctarr] (xt2) to  (lst2);
\draw [fctarr] (xts) to  (lsts);

\draw [fctarr] (lst) to  (ht);
\draw [fctarr] (lst1) to  (ht1);
\draw [fctarr] (ht1) to  (lst);
\draw [fctarr] (lst2) to  (ht2);
\draw [fctarr] (ht2) to  (lst1);
\draw [fctarr] (lst) to  (ct);
\draw [fctarr] (lst1) to  (ct1);
\draw [fctarr] (ct1) to  (lst);
\draw [fctarr] (lst2) to  (ct2);
\draw [fctarr] (ct2) to  (lst1);

\draw [fctarr] (lspt) to  (hpt);
\draw [fctarr] (lspt1) to  (hpt1);
\draw [fctarr] (hpt1) to  (lspt);
\draw [fctarr] (lspt2) to  (hpt2);
\draw [fctarr] (hpt2) to  (lspt1);
\draw [fctarr] (lspt) to  (cpt);
\draw [fctarr] (lspt1) to  (cpt1);
\draw [fctarr] (cpt1) to  (lspt);
\draw [fctarr] (lspt2) to  (cpt2);
\draw [fctarr] (cpt2) to  (lspt1);

\draw [fctarr, dotted] (ht) to  [in=270, out=135] (lspt); %
\draw [fctarr, dotted] (ht1) to  [in=270, out=135] (lspt1);
\draw [fctarr, dotted] (ht2) to  [in=270, out=135] (lspt2);

\draw [fctarr] (hpt) to  (out);

\node[dot] at (-2.5*\lstmsep-2mm,0) {\pgfuseplotmark{*}};
\node[dot] at (-2.5*\lstmsep,0) {\pgfuseplotmark{*}};
\node[dot] at (-2.5*\lstmsep+2mm,0) {\pgfuseplotmark{*}};
\node[dot] at (-2.5*\lstmsep-2mm,\stepsep) {\pgfuseplotmark{*}};
\node[dot] at (-2.5*\lstmsep,\stepsep) {\pgfuseplotmark{*}};
\node[dot] at (-2.5*\lstmsep+2mm,\stepsep) {\pgfuseplotmark{*}};
\node[dot] at (-2.5*\lstmsep-2mm,2*\stepsep) {\pgfuseplotmark{*}};
\node[dot] at (-2.5*\lstmsep,2*\stepsep) {\pgfuseplotmark{*}};
\node[dot] at (-2.5*\lstmsep+2mm,2*\stepsep) {\pgfuseplotmark{*}};
\node[dot] at (-3*\lstmsep,1.5*\stepsep+1.5mm) {\pgfuseplotmark{*}};
\node[dot] at (-3*\lstmsep,1.5*\stepsep) {\pgfuseplotmark{*}};
\node[dot] at (-3*\lstmsep,1.5*\stepsep-1.5mm) {\pgfuseplotmark{*}};
\draw [usearr] (out) to (L);
\draw [usearr] (y) to (L);
\end{tikzpicture}
\caption{Multilayer LSTM network flowchart from input $\tilde{\bm{x}}_t:=(\bm{x}_{t-s},\ldots,\bm{x}_{t-1})$ to output $\tilde{g}_\mathcal{W}(\tilde{\bm{x}}_t)$, with loss evaluation. The LSTM cells represent the transformation in~\eqref{lstm_cell}.}%
\label{f:MLLSTMdiagr}
\end{figure}

\section{Simulation study for independent observations}\label{ss:simsiid}

Three data-generating models with independent observations are considered. The training sample $\mathcal{D}=\left\{(\bm{x}_i, y_i)\right\}_{i=1}^n$ is drawn from
\begin{equation}
\begin{cases}
\bm{X}  \sim \mathcal{U}\left([-1, 1]^{p}\right),\\
Y \mid \bm{X} = \bm{x} \sim \sigma(\bm{x})\cdot t_{\alpha(\bm{x})},
\end{cases}
\end{equation}
with $p=10$, $\alpha(\bm{x}) = 1/\xi(\bm{x}): = 7\cdot\{1 + \exp(4x_1 + 1.2)\}^{-1} + 3$, and three different models for~$\sigma(\bm{x})$:
\begin{enumerate}
\item[] Model 1: $\sigma(\bm{x}): = 1 + 6 \phi(x_1, x_2)$, where $\phi$ is the bivariate Gaussian density with correlation $0.9$,
\item[] Model 2: $\sigma(\bm{x}): = 4 + 3\cos(7\left\Vert(x_{1},x_{2})^\top\right\Vert_2+3)$,
\item[] Model 3: $\sigma(\bm{x}): = 4 + 3\cos(6\left\Vert\bm{x}\right\Vert_2+3.5)$.
\end{enumerate}
To avoid bias in the selection of the data models, $\alpha(\bm{x})$ and $\sigma(\bm{x})$ in Model~1 are the same as in the simulation study in~\citet{gbex}. The choices for Models~2 and~3 are designed to study more complex covariate dependencies. The constants are chosen to have positive scale values and enough variation for the inference task to be interesting.

As the focus of the experiments is on the extremal part of the model and since all extreme value competitors use the same intermediate quantile estimates, we use the true $Q_{\bm{x}}(\tau_0)$ as intermediate quantiles.
Two main types of network architectures are considered for EQRN. The first type is an MLP with $\tanh$ activation functions, narrow architectures with between one and four hidden layers and optional $L_2$ weight penalty as training regularization. The second type is self-normalizing networks~\citep{SNN} using SELU activation functions, deeper architectures with between three and eight hidden layers and optional alpha dropout as regularization. This type of network is designed to maintain unit variance and zero mean across layers in deeper networks trained for regression tasks, in order to avoid vanishing and exploding gradient issues. Results suggest that the additional flexibility of the second type was not necessary for the three tasks at hand, as they never yielded better validation scores than the best $\tanh$ models.
Table~\ref{t:simstructures} summarizes the hyperparameters of the chosen EQRN networks.

To evaluate the accuracy of the best models over the full feature space $\mathcal{X}=[-1,1]^p$, we use the integrated squared error (ISE) between the prediction $\hat{Q}_\bm{x}(\tau)$ and the true quantile $Q_\bm{x}(\tau)$,
\begin{equation}
    \int_\mathcal{X} \left(\hat{Q}_\bm{x}(\tau) - Q_\bm{x}(\tau) \right)^2 d\bm{x}.
\end{equation}
We generate test features using a Halton sequence~\citep{halton1964} and compute the MSE between the corresponding predicted and true response quantiles, to estimate the $p$-dimensional integral.

\begin{figure}[t]
\centering
\includegraphics[width=\textwidth]{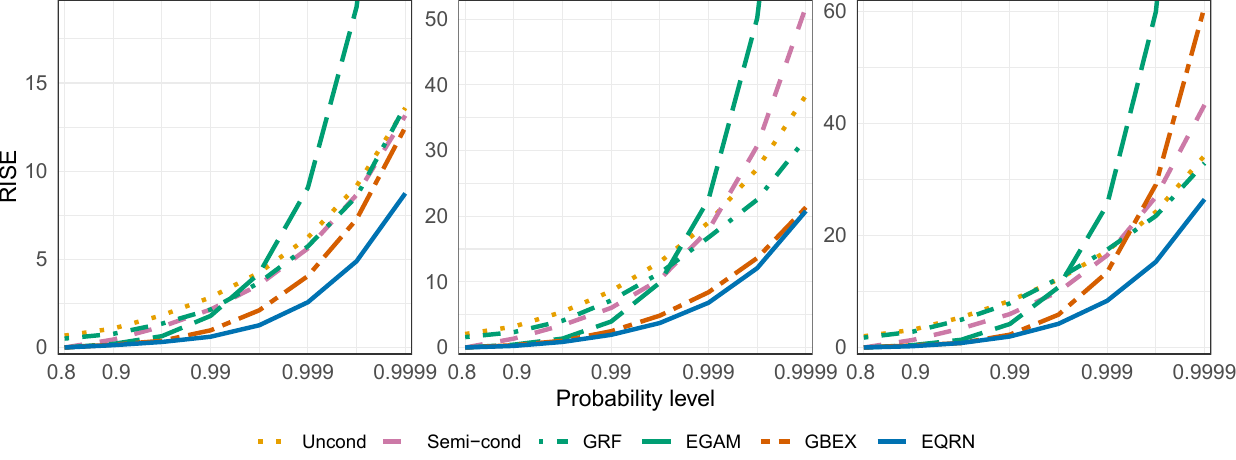}%
\caption{Root integrated squared error between predicted and true conditional quantiles at different probability levels $\tau$ (log-scale) for the selected EQRN model and the improved competitors, for data Models 1--3 (left to right). The cropped-out RMISE for EGAM at level 0.9999 are around 43, 115 and 150, respectively.}
\label{f:indisecomp}
\end{figure}

Figure~\ref{f:indisecomp} shows the accuracy of EQRN and the competitor methods for an increasingly large $\tau$, for the three data models. The competitors' performances shown here were also significantly improved by using the intermediate $\hat{Q}_{{\bm{x}}}(\tau_0)$ as an additional covariate. 
For every model, EQRN outperforms all competitors, with a difference in accuracy increasing with $\tau$. EGAM seems to suffer from the large dimension of the feature space at large $\tau$, both when the actual quantile depends on only two or all of the 10 features. The difference in accuracy of EQRN and GBEX compared to the unconditional and semiconditional models is particularly significant for Models~1 and~3, and EQRN generally outperforms GBEX, especially at high probability levels.

\begin{table}[t]
    \caption{Hyperparameters of the EQRN networks with the best validation loss for the three independent data models.}%
    \label{t:simstructures}
    \centering
    \begin{tabular}{l|ccc}%
         \hline
         & Hidden activation function & Hidden layer dimensions & $L_2$ penalty \\
         \hline
         Model~1 & $\tanh$ & $(128,128,128)$ & $10^{-5}$ \\
         Model~2 & $\tanh$ & $(20,10,10)$ & $10^{-5}$ \\
         Model~3 & $\tanh$ & $(10,10,10)$ & $0$ \\
         \hline
    \end{tabular}%
\end{table}

We also define the quantile R squared of $\hat{Q}_\bm{x}(\tau)$ over the sample $\mathcal{D}$ as
\begin{equation}\label{e:R2quant}
    R^2_\tau := 1 - \dfrac{\sum_{i=1}^n\left(Q_{\bm{x}_i}(\tau) - \hat{Q}_{\bm{x}_i}(\tau) \right)^2}{\sum_{i=1}^n\left(Q_{\bm{x}_i}(\tau) - \overline{Q_\mathcal{D}(\tau)}\right)^2}, \quad \text{ with }\quad \overline{Q_\mathcal{D}(\tau)} := \dfrac{1}{n}\sum_{i=1}^n Q_{\bm{x}_i}(\tau).
\end{equation}
The definition is similar to the classical R squared coefficient of determination in regression, but the true conditional quantile values are used as targets instead of the response observations. The $R^2_\tau$ is essentially the reversed MSE normalized by the variance of the true conditional quantile. A value close to unity indicates a very low MSE compared to the quantile variance, and negative values indicate a MSE larger than the quantile variance.

\begin{figure}[t]
\centering
\includegraphics[width=\textwidth]{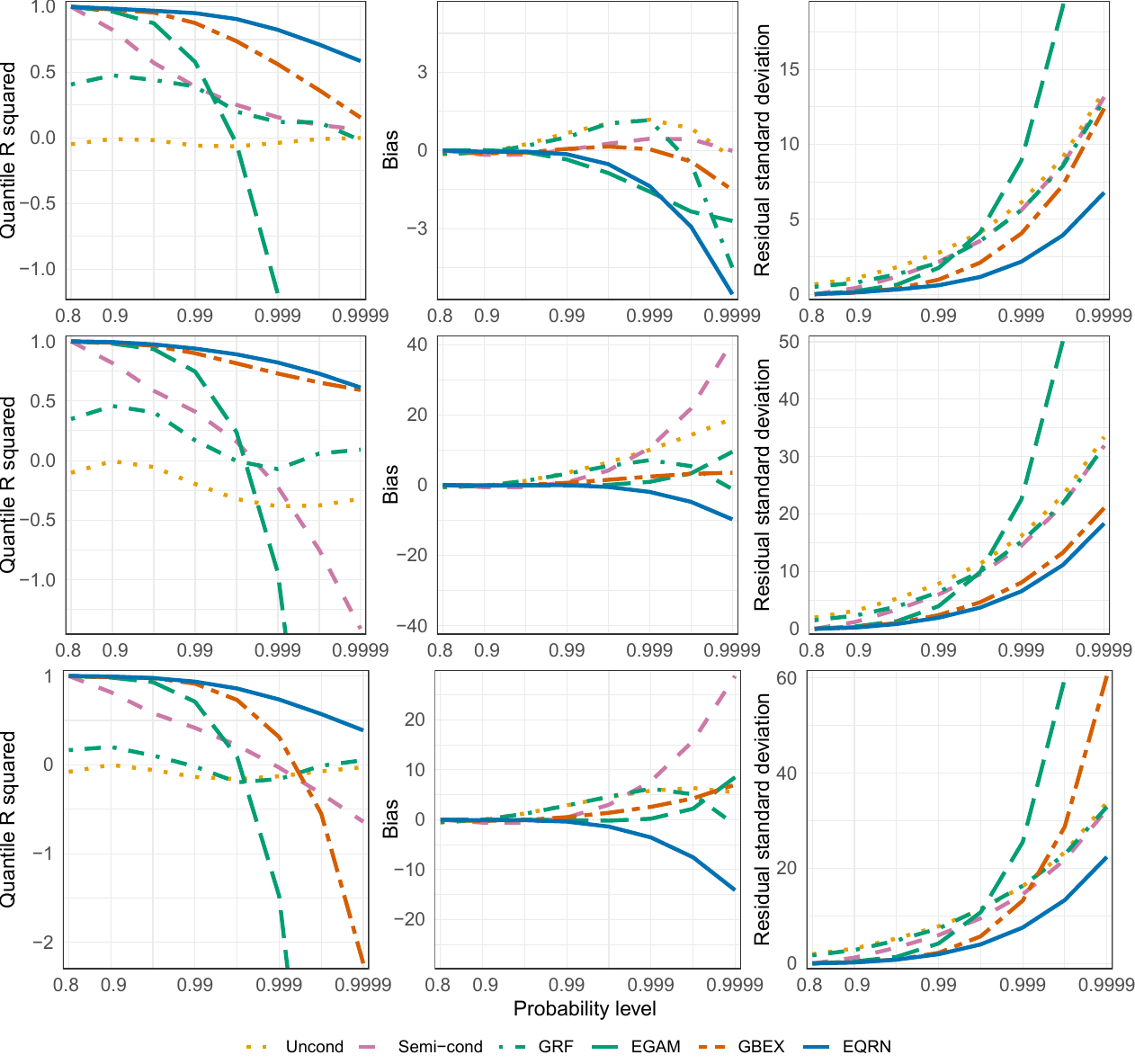}
\caption{Quantile R squared, bias and residual standard deviation of the predicted quantiles compared to the truth at different probability levels (log-scale) for the selected EQRN model and improved competitors, for data Models~1--3 (top to bottom).}
\label{f:indRbvcomp}
\end{figure}

Figure~\ref{f:indRbvcomp} shows the quantile R squared, the biases and the residual standard deviations of the same respective quantile predictions compared to the truth. The R squared lead to the same conclusions as the RISE. Regarding the bias-variance decomposition of the RISE, it seems that the variance term is dominating the square bias. Although EQRN is here not the least biased model for large $\tau$ values, it is its dominating performance in terms of residual variance that leads to it having the lowest RISE values for every data model.

\begin{figure}[t]
\centering
\includegraphics[width=0.9\textwidth]{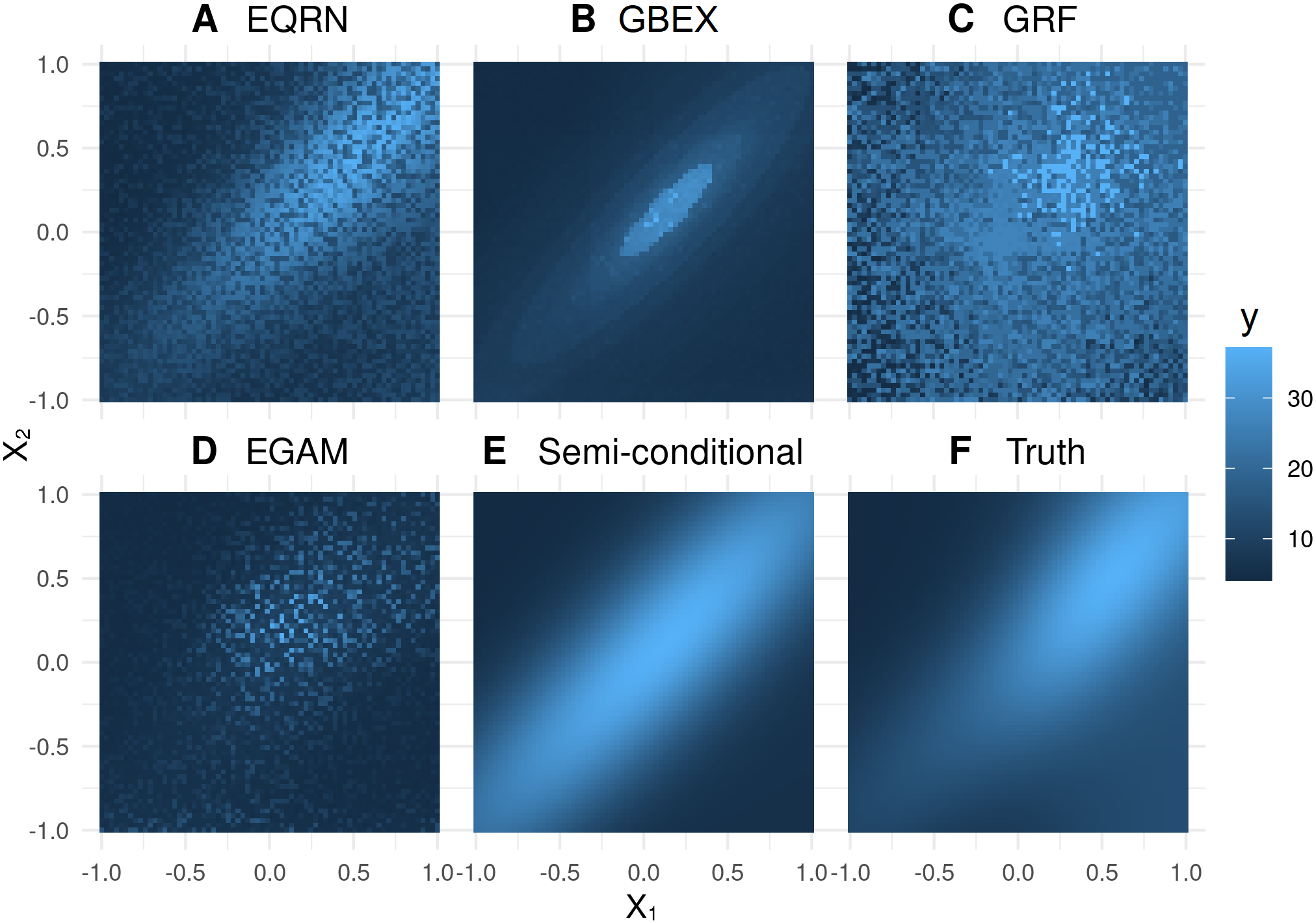}%
\caption{Conditional quantile predictions of EQRN and the improved competitor models at probability level $\tau=0.9995$, shown as a function of $X_1$ and $X_2$, for Model~1.}
\label{f:predsbinorm}
\end{figure}

Figure~\ref{f:predsbinorm} shows the predicted $\hat{Q}_\bm{x}(0.9995)$ for EQRN and the competitors, as a function of the two significant covariates $(x_1,x_2)$ for Model~1.
At that extreme level, EQRN still seems to capture the true conditional quantile function quite well, although the predictions show some residual noise.
GBEX shows an elliptical stepwise approximation behaviour. and seems to underestimate the smaller quantiles and overestimate the largest quantiles.
EGAM and GRF fail drastically at recovering the conditional quantile function. The semiconditional estimates are a translation of the intermediate quantiles, which in particular fail to capture the varying shape along $X_1$.

\newpage

\section{Simulation study for sequentially dependent data}\label{ss:simsseqsuppl}

\begin{figure}[t]
\centering
\includegraphics[width=0.85\textwidth]{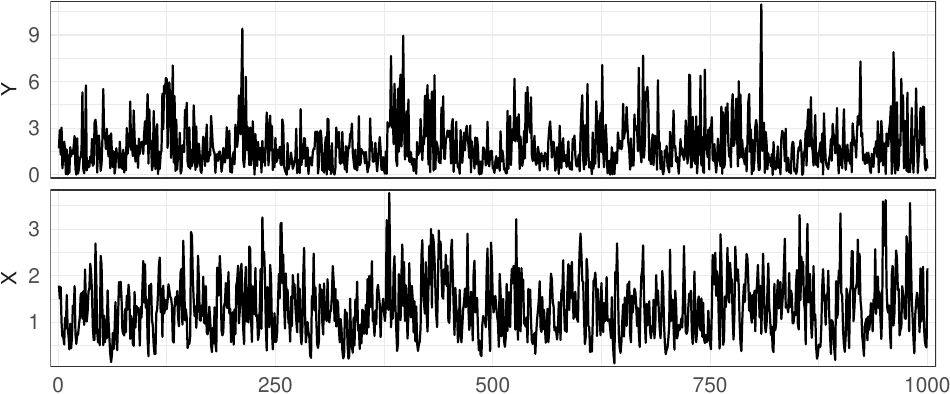}%
\caption{First $\numprint{1000}$ observations of the sequential data simulated from~\eqref{e:tssim}.}%
\label{f:tsdata}
\end{figure}

The main results of the simulation study on sequentially dependent data are presented in the main paper. This section discusses additional results. Figure~\ref{f:tsdata} shows part of the sequential data simulated from the generating process described in the main paper.

\begin{figure}[t]
\centering
\includegraphics[width=\textwidth]{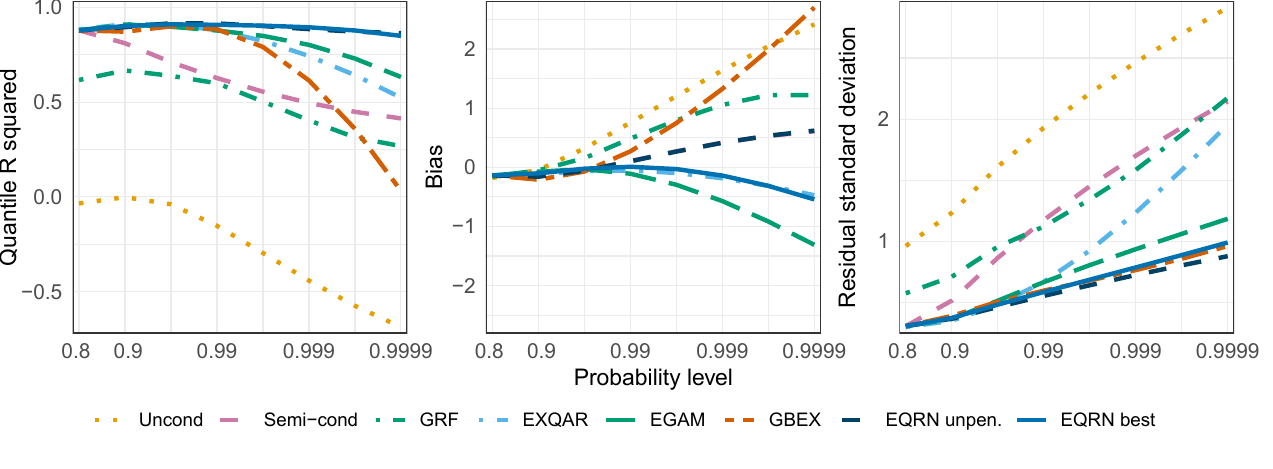}%
\caption{Quantile R squared, bias and residual standard deviation of the predicted quantiles compared to the truth at different probability levels (log-scale) for the selected EQRN model and the improved competitors, for the sequential data model.}
\label{f:seqRbvcomp}
\end{figure}

Figure~\ref{f:seqRbvcomp} shows the quantile R squared~\eqref{e:R2quant}, the biases and the residual standard deviations of the quantile predictions compared to the truth, for the two selected EQRN models and competitors. The R squared evolution again shows EQRN is the model that best captures the covariate sequential dependence in the tail, as it outperforms all competitors with a difference in accuracy increasing with $\tau$. In terms of bias, the penalized EQRN here scores similar values as EXQAR, and both EQRN versions outperform all other methods. The unpenalized EQRN has the lowest residual variance, closely followed by both GBEX and the penalized EQRN, although GBEX has a bad accuracy overall, due to its large bias. 

\begin{figure}[t]
\centering
\includegraphics[width=\textwidth]{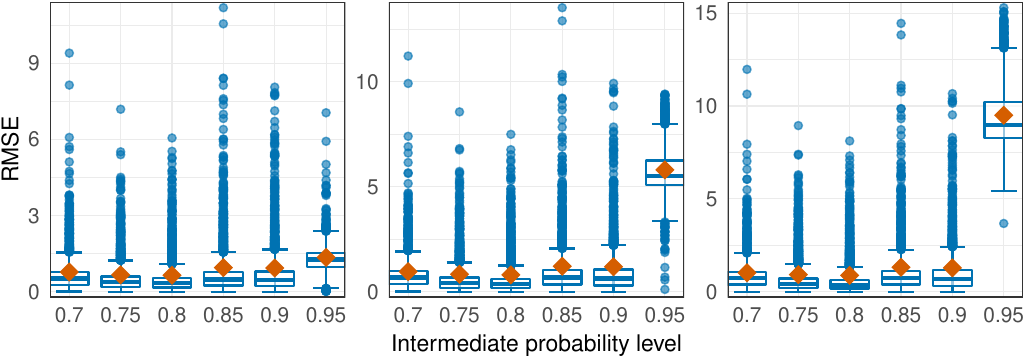}%
\caption{Boxplot of the absolute residuals of the quantile predictions from the selected EQRN model (blue) and their RMSE (red diamond) at probability levels $0.995$ (left), $0.999$ (middle) and $0.9995$ (right) for different choices of intermediate probability level $\tau_0$, for the sequential data model.}
\label{f:seqthreshbox}
\end{figure}

Figure~\ref{f:seqthreshbox} shows the impact of the intermediate level $\tau_0$ on the accuracy of the best EQRN model. The value $\tau_0=0.8$ used in the rest of the analysis leads to the best RMSE. This relatively low value shows that $\tau_0$ can be chosen much lower than for the classical unconditional GPD model. An intuitive explanation for this fact is the following. In a situation without covariates, the choice of $\tau_0$ is a trade-off between approximation bias (which favours larger thresholds) and variance (which favours lower thresholds); see Figure 3 in the main document. For covariate-dependent data, the distribution of the exceedances varies and more data is needed to accurately capture this function of the covariates. The variance becomes more important than the approximation bias, therefore, lower thresholds are preferable. 
Moreover, the flexibility of the GPD regression neural network model seems to be able to absorb some of the approximation bias, also allowing for a low value of $\tau_0$.

The final accuracy seems in fact to not be too sensitive to the choice for $\tau_0$ compared to the network's grid-searched hyperparameters, discussed in the main analysis, as the differences in RMSE for $\tau_0$ values close to the optimum are relatively small. 
As mentioned in Section~\ref{s:eqrnn} of the main paper, $\tau_0$ cannot be treated as a classical tuning parameter, as different values for $\tau_0$ generally yield different subsets of exceedances. Thus, the likelihood~\eqref{e:ogpdlik}, which is used as a goodness of fit metric for hyperparameter tuning, would not be comparable between models. The RMSE cannot be computed in practice either, since $Q_\bm{x}(\tau)$ is generally unknown.

\section{Application: competitor results}\label{ss:applsuppl}

The main results from our application to forecasting flood risk in Switzerland using our proposed EQRN methodology are presented in the main paper. This section discusses and compares additional results using the competitor methods, adapted to provide the same type of forecast as the EQRN approach, with a focus on the 2005 flood event (see Figure~\ref{f:chpreds2005} in the main paper).

\begin{figure}[t]
\centering
\includegraphics[width=0.9\textwidth]{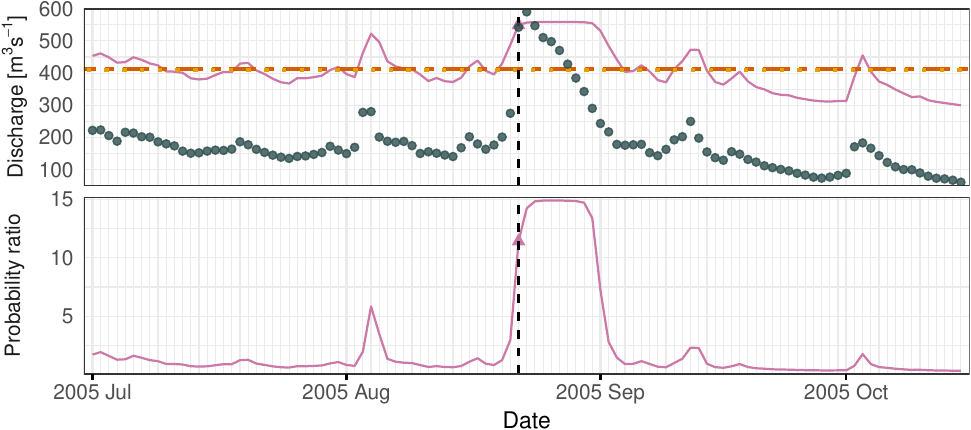}%
\caption{Top: Daily average discharge (points) at the Bern--Sch\"onau station (62) 
and one-day-ahead semiconditional forecasts of conditional 100-year quantiles (solid line) during the 2005 flood. Horizontal lines show unconditional $Q^{100}$ based on GEV (dashed) and GPD (dotted). Bottom: One-day-ahead forecast of the conditional probability of exceeding the GEV estimated $Q^{100}$ as a ratio to the unconditional probability, using the semiconditional parameter forecast. The vertical line indicates August 22, the day of the first exceedance.}
\label{f:CHsemicond}
\end{figure}

\begin{figure}[t]
\centering
\includegraphics[width=0.9\textwidth]{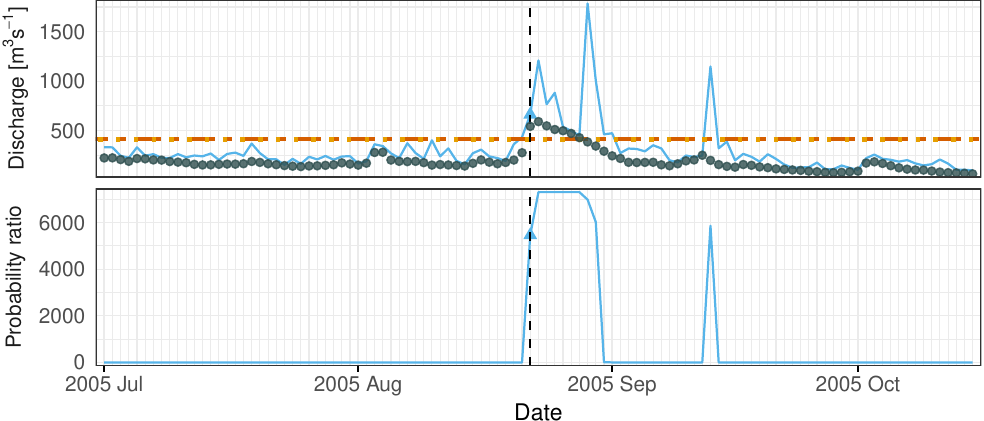}%
\caption{Top: Daily average discharge (points) at the Bern--Sch\"onau station (62) 
and one-day-ahead EXQAR forecasts of conditional 100-year quantiles (solid line) during the 2005 flood. Horizontal lines show unconditional $Q^{100}$ based on GEV (dashed) and GPD (dotted). Bottom: One-day-ahead forecast of the conditional probability of exceeding the GEV estimated $Q^{100}$ as a ratio to the unconditional probability, using the EXQAR forecast. The vertical line indicates August 22, the day of the first exceedance.}
\label{f:CHexqar}
\end{figure}

\begin{figure}[t]
\centering
\includegraphics[width=0.9\textwidth]{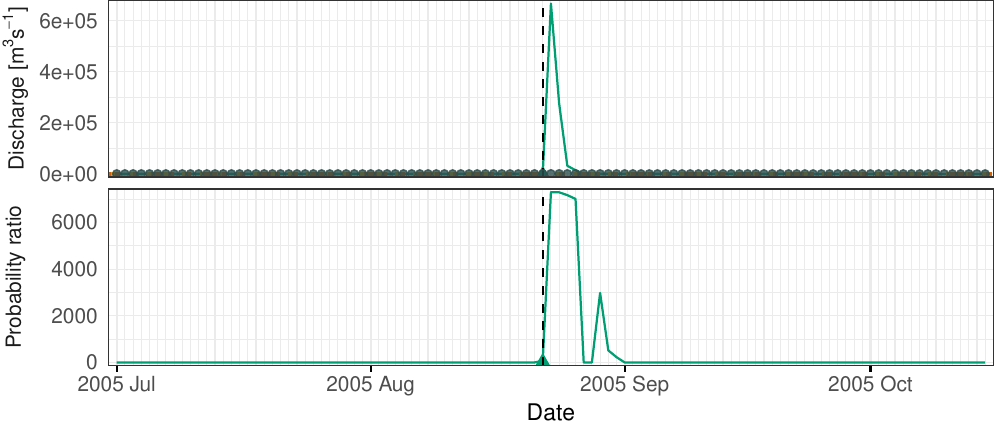}%
\caption{Top: Daily average discharge (points) at the Bern--Sch\"onau station (62) 
and one-day-ahead EGAM forecasts of conditional 100-year quantiles (solid line) during the 2005 flood. Horizontal lines show unconditional $Q^{100}$ based on GEV (dashed) and GPD (dotted). Bottom: One-day-ahead forecast of the conditional probability of exceeding the GEV estimated $Q^{100}$ as a ratio to the unconditional probability, using the EGAM parameter forecast. The vertical line indicates August 22, the day of the first exceedance.}
\label{f:CHegam}
\end{figure}

\begin{figure}[t]
\centering
\includegraphics[width=0.9\textwidth]{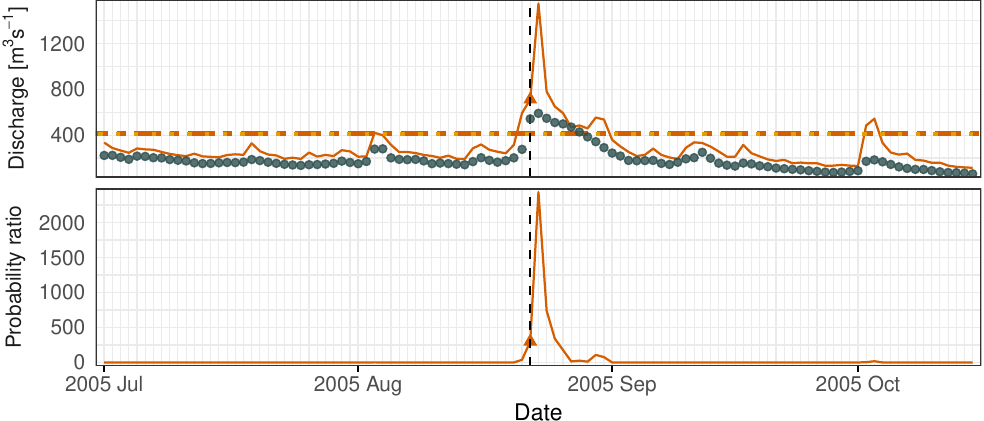}%
\caption{Top: Daily average discharge (points) at the Bern--Sch\"onau station (62) 
and one-day-ahead GBEX forecasts of conditional 100-year quantiles (solid line) during the 2005 flood. Horizontal lines show unconditional $Q^{100}$ based on GEV (dashed) and GPD (dotted). Bottom: One-day-ahead forecast of the conditional probability of exceeding the GEV estimated $Q^{100}$ as a ratio to the unconditional probability, using the GBEX parameter forecast. The vertical line indicates August 22, the day of the first exceedance.}
\label{f:CHgbex}
\end{figure}

We first observe that the predictions have roughly the same behaviour, which shows that all methods capture at least some of the temporal structure based on the past covariates. The semiconditional method (Figure~\ref{f:CHsemicond}) is clearly not flexible enough, since the predictions only show a weak sensitivity to the changes in covariates. It also fails to trigger any early warning during the main event, due to low probability ratio forecasts never exceeding the selected threshold value of $100$. The reason is that, while the intermediate quantile is covariate-dependent, the GPD parameters are constant over time. We conclude that a covariate-dependent model for the tail is required in this application.    
Figure~\ref{f:CHexqar} shows predictions from the EXQAR model~\citep{EXQAR}. This model is more sensitive to changes in the covariates, but the regression function looks fairly erratic, with sudden spikes at some time points. Those spikes are here due to unusually small shape estimates in combination with a large-scale estimate. This might be caused by an instability in the estimation of the local moments used in the model. EGAM (Figure~\ref{f:CHegam}) fails to trigger an early warning for the first day of the flooding event as its quantile and probability ratio forecasts are very low, but then seems to severely overestimate the river flow during the rest of the event.
The best competing model seems to be the GBEX (Figure~\ref{f:CHgbex}), as it yields a smooth prediction curve with a pronounced spike at the main event. Comparing this with our EQRN method (Figure~2 in the main paper), it reacts slightly later and its risk forecast (probability ratio) on the day before the first exceedance is significantly lower than for EQRN.

\begin{figure}[t]
\centering
\includegraphics[width=0.5\textwidth]{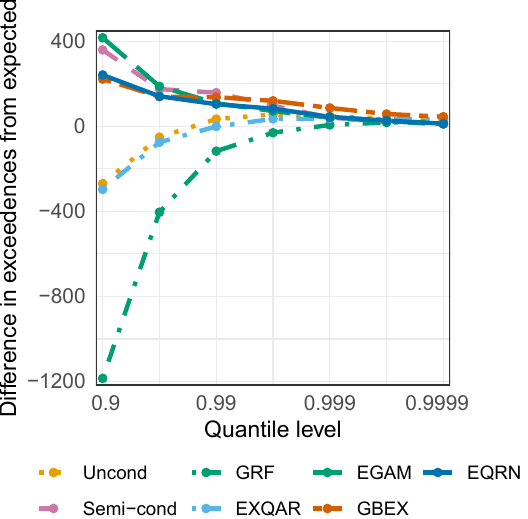}%
\caption{Difference between the number of observations exceeding the EQRN and competitor quantile predictions on the test set and the expected number of exceedances, for different probability levels (log-scale).}%
\label{f:compcalibr}
\end{figure}

This qualitative way of checking the model is important since in real-world applications, the true extreme quantiles are unknown. Therefore, only some quantitative model checks can be performed, like the right-hand panel of Figure~\ref{f:chnonstaticalibr} in the main paper showing calibration in the number of quantile-exceeding test observations. Figure~\ref{f:compcalibr} compares this number of quantile-exceedances for the competitor predictions. Although they can give evidence against the suitability of a model, it highlights that such metrics only assess calibration but not goodness-of-fit nor accuracy, as unconditional methods obtain similar values to flexible accurate methods. 
This underlines the importance of the simulation studies, which allow us to evaluate in several settings which methods are more accurate. In particular, in situations with temporal dependence, our EQRN method outperforms the competitors (e.g., Figure~\ref{f:tscompall} in the main paper).

\newpage

\end{appendix}

\end{document}